\documentclass[preprint,review,12pt]{elsarticle}
 
\usepackage{graphics}
\usepackage{graphicx}

\usepackage{epsfig}

\usepackage{amssymb}
\usepackage{amsthm}

\theoremstyle{plain}

\theoremstyle{definition}

\theoremstyle{remark}

\usepackage{lineno}
\usepackage{color}

\begin{document}

\begin{frontmatter}

 

\title{
Liouville Spectral Gap and Bifurcation--Driven Lagrangian--Eulerian Decoupling with Nondiffusive Turbulence Closures
}
 

\author{Nicola de Divitiis}

\address{"La Sapienza" University, Dipartimento di Ingegneria Meccanica e Aerospaziale, 
Via Eudossiana, 18, 00184 Rome, Italy, \\
phone: +39--0644585268, \ \ fax: +39--0644585750, \\ 
e-mail: n.dedivitiis@gmail.com, \ \  nicola.dedivitiis@uniroma1.it}

\begin{abstract}
In fully developed homogeneous and isotropic turbulence,  the Lagrangian and Eulerian descriptions of motion, although formally equivalent, become statistically decoupled. In this work, by invoking Liouville theorem, we show that the joint probability density function (PDF) of the Eulerian and Lagrangian fields, evolving from arbitrary initial conditions, relaxes exponentially toward a factorized form given by the product of the corresponding marginal PDFs. This relaxation is governed by a genuine spectral gap of the Liouville operator, whose magnitude is primarily set by the bifurcation rate of the velocity-–gradient dynamics, whereas the contribution of Lyapunov exponents is shown to be significantly smaller.
As a consequence, Eulerian-–Lagrangian correlations decay rapidly, and if the joint PDF is initially factorized, its factorized structure is preserved at all subsequent times, with each marginal evolving independently under the corresponding dynamics. We further show that the formal equivalence between the two descriptions implies the invariance of the relative kinetic energy between arbitrarily chosen points. When combined with the asymmetric statistics of instantaneous finite-–scale Lyapunov exponents in incompressible turbulence, this property provides a quantitative interpretation of particle-–pair separation and of the turbulent energy cascade.

Finally, these results naturally lead to nondiffusive closure relations for the von K\'arm\'an-–Howarth and Corrsin equations, which coincide with those previously proposed by the author, thereby providing an independent theoretical validation of those closures.
\end{abstract}

\vspace{5.mm}


\end{frontmatter}

\newcommand{\tr}{ \mbox{tr} }
\newcommand{\no}{\noindent}
\newcommand{\be}{\begin{equation}}
\newcommand{\ee}{\end{equation}}
\newcommand{\bea}{\begin{eqnarray}}
\newcommand{\eea}{\end{eqnarray}}
\newcommand{\bc}{\begin{center}}
\newcommand{\ec}{\end{center}}

\newcommand{\calr}{{\cal R}}
\newcommand{\calv}{{\cal V}}

\newcommand{\bff}{\mbox{\boldmath $f$}}
\newcommand{\bfg}{\mbox{\boldmath $g$}}
\newcommand{\bfh}{\mbox{\boldmath $h$}}
\newcommand{\bfi}{\mbox{\boldmath $i$}}
\newcommand{\bfm}{\mbox{\boldmath $m$}}
\newcommand{\bfp}{\mbox{\boldmath $p$}}
\newcommand{\bfr}{\mbox{\boldmath $r$}}
\newcommand{\bfu}{\mbox{\boldmath $u$}}
\newcommand{\bfv}{\mbox{\boldmath $v$}}
\newcommand{\bfx}{\mbox{\boldmath $x$}}
\newcommand{\bfy}{\mbox{\boldmath $y$}}
\newcommand{\bfw}{\mbox{\boldmath $w$}}
\newcommand{\bfk}{\mbox{\boldmath $\kappa$}}

\newcommand{\bfA}{\mbox{\boldmath $A$}}
\newcommand{\bfD}{\mbox{\boldmath $D$}}
\newcommand{\bfI}{\mbox{\boldmath $I$}}
\newcommand{\bfL}{\mbox{\boldmath $L$}}
\newcommand{\bfM}
{\mbox{\boldmath $M$}}
\newcommand{\bfS}{\mbox{\boldmath $S$}}
\newcommand{\bfT}{\mbox{\boldmath $T$}}
\newcommand{\bfU}{\mbox{\boldmath $U$}}
\newcommand{\bfX}{\mbox{\boldmath $X$}}
\newcommand{\bfY}{\mbox{\boldmath $Y$}}
\newcommand{\bfK}{\mbox{\boldmatthe average of $u_\xi u_\xi^*/u^2$h $K$}}

\newcommand{\bfeta}{\mbox{\boldmath $\eta$}}
\newcommand{\bfiota}{\mbox{\boldmath $\iota$}}
\newcommand{\bfrho}{\mbox{\boldmath $\rho$}}
\newcommand{\bfchi}{\mbox{\boldmath $\chi$}}
\newcommand{\bfphi}{\mbox{\boldmath $\phi$}}
\newcommand{\bfPhi}{\mbox{\boldmath $\Phi$}}
\newcommand{\bflambda}{\mbox{\boldmath $\lambda$}}
\newcommand{\bfxi}{\mbox{\boldmath $\xi$}}
\newcommand{\bfLambda}{\mbox{\boldmath $\Lambda$}}
\newcommand{\bfPsi}{\mbox{\boldmath $\Psi$}}
\newcommand{\bfomega}{\mbox{\boldmath $\omega$}}
\newcommand{\bfOmega}{\mbox{\boldmath $\Omega$}}
\newcommand{\bfeps}{\mbox{\boldmath $\varepsilon$}}
\newcommand{\bfepsn}{\mbox{\boldmath $\epsilon$}}
\newcommand{\bfzeta}{\mbox{\boldmath $\zeta$}}
\newcommand{\bfkappa}{\mbox{\boldmath $\kappa$}}
\newcommand{\bfsigma}{\mbox{\boldmath $\sigma$}}
\newcommand{\bftau}{\mbox{\boldmath $\tau$}}
\newcommand{\itPsi}{\mbox{\it $\Psi$}}
\newcommand{\itPhi}{\mbox{\it $\Phi$}}

\newcommand{\bint}{\mbox{ \int{a}{b}} }

\newcommand{\ds}{\displaystyle}
\newcommand{\Sum}{\Large \sum}



\bigskip

\section{Introduction \label{intro}}

Classical studies of homogeneous isotropic turbulence have been predominantly formulated within the Eulerian description of fluid motion, leading to evolution equations for velocity and temperature correlations defined over ensembles of spatial fields \cite{Karman38, Corrsin_1, Corrsin_2}. While this approach provides a systematic framework for the statistical analysis of turbulence, it does not yield a fundamental explanation of the energy cascade, nor does it supply closure relations for the convective terms of the correlation equations, except under specific modeling assumptions \cite{Hasselmann58, Millionshtchikov69, Oberlack93, Baev, Mellor84}. In particular, the dynamics of fluid particle displacements and trajectories, which play a central role in turbulent transport and mixing, are not explicitly represented at the level of Eulerian correlation equations.

This limitation can be traced to the fact that, although particle transport is implicitly embedded in the Eulerian formulation, it is not explicitly resolved in terms of the evolution of fluid displacements. By contrast, such quantities are naturally defined within the Lagrangian description of motion. Since the displacement of a mechanical system contributes to the definition of its state of motion, whose phase space is spanned by generalized coordinates and velocities, a reduction of the fluid–-dynamical phase space to Eulerian fields alone may lead to an incomplete representation of the mechanisms governing turbulence, at least within the framework of correlation equations.

Under very general smoothness conditions, however, the Eulerian and Lagrangian descriptions constitute formally equivalent representations of fluid motion \cite{Truesdell77}. In the framework of this equivalence, the central objective of the present work is to exploit a specific spectral analysis of the Liouville operator together with the finite--scale Lyapunov theory based on bifurcations, to investigate the statistical correlation between Eulerian and Lagrangian fields. In particular, we show that the invariance of the relative kinetic energy between two points, when expressed in either representation of motion, provides a fundamental mechanism promoting particle trajectory separation and the turbulent energy cascade.

To this end, the concept of bifurcation rate is introduced as a key quantity characterizing the chaotic dynamics of turbulent flows. The bifurcation rate is defined as the average frequency at which bifurcations occur during chaotic regimes and corresponds to the rate at which the trajectories intersect $\Sigma_D$, the hypersurface of phase space on which the Jacobian of the dynamical system becomes singular. This quantity thus represents the mean crossing rate of the Jacobian--degeneracy manifold. If trajectories do not intersect $\Sigma_D$, the system although nonlinear—does not exhibit chaotic behavior. Conversely, when trajectories repeatedly intersect $\Sigma_D$, chaotic dynamics arise and the state variables fluctuate on timescales determined by the bifurcation rate. In the present analysis, two distinct bifurcation rates are considered: one associated with the Navier--Stokes equations (Eulerian bifurcations) and one associated with the velocity gradient (Lagrangian bifurcations).

Although a large body of literature has addressed homogeneous isotropic turbulence and the closure problem of correlation equations \cite{Hasselmann58, Millionshtchikov69, Oberlack93, Baev, Mellor84, George1, George2, Antonia, Onufriev94, Grebenev05, Grebenev09, Antonia2013}, a unified treatment combining Liouville spectral theory, bifurcation dynamics,  and finite--scale Lyapunov analysis has, to the author’s knowledge, not been previously presented. 
The aim of the present work is therefore to employ this combined framework to study the decay of Eulerian-–Lagrangian correlations and to demonstrate that the effect of the lagrangian bifurcation rate
leads naturally to a statistical decoupling of Eulerian and Lagrangian fields, whereas
the equivalence between the two descriptions of motion gives a physically grounded interpretation of the turbulent energy cascade.

The paper first recalls the elements of continuum kinematics relevant to the present analysis \cite{Truesdell77}. The Navier--Stokes and heat equations, together with the displacement equation, are then cast into a symbolic operator form to formulate Liouville theorem for both Eulerian and Lagrangian fields. This formulation allows Navier--Stokes bifurcations and Lagrangian bifurcations to be defined in close analogy with ordinary differential equations and enables the investigation of several properties of fully developed turbulence, some of which have been examined previously by the author \cite{deDivitiis_1, deDivitiis_4, deDivitiis_5, deDivitiis_8, deDivitiis_9, deDivitiis_10}. The novel contributions of the present work are summarized as follows.

(i) The order of magnitude of the Navier--Stokes bifurcation rate is estimated from the mathematical structure of the Navier--Stokes equations and from the decay properties of homogeneous isotropic turbulence. In addition, the bifurcation rate of the velocity gradient is estimated on the basis of the particular mathematical structure of the Eulerian velocity field which in turn is the result of the
Navier--Stokes bifurcations.

(ii) A specific Liouville spectral analysis is presented which relates Liouville eigenvalues,  bifurcation rates, and Lyapunov exponents, and an interpretation of such link in the framework of Ruelle--Pollicott resonances \cite{Pollicott85, Ruelle86, Ruelle87} is given, with particular reference to the combined effect of bifurcation rate and Lyapunov exponents on Liouville eigenvalues.

(iii) It is shown that the joint probability density function (PDF) of the Eulerian and Lagrangian fields, evolving from arbitrary initial conditions, relaxes exponentially toward a factorized form given by the product of the corresponding marginal PDFs. This relaxation is governed by a genuine spectral gap of the Liouville operator, whose magnitude is primarily determined by the bifurcation rate of the velocity--gradient dynamics, while the contribution of Lyapunov exponents is comparatively small. As a consequence, Eulerian–-Lagrangian correlations decay rapidly, and if the joint PDF is initially factorized, this structure is preserved at all subsequent times, with each marginal evolving independently under the corresponding dynamics.

(iv) The formal equivalence between the Eulerian and Lagrangian descriptions implies the invariance of the relative kinetic energy between arbitrarily chosen points. When combined with the asymmetric statistics of instantaneous finite--scale Lyapunov exponents in incompressible turbulence, this property provides a quantitative interpretation of particle–-pair separation and of the turbulent energy cascade.

(v) A detailed finite--scale Lyapunov analysis based on the preceding results is presented, leading to estimates of the range of finite--scale Lyapunov exponents of the velocity gradient and of their probability distribution.

Finally, closure relations for the von K\'arm\'an--Howarth and Corrsin equations are derived using Liouville theorem and the associated spectral gap. These closures coincide with those previously obtained by the author \cite{deDivitiis_1, deDivitiis_4, deDivitiis_5, deDivitiis_8, deDivitiis_9, deDivitiis_10}. The resulting nondiffusive formulas describe a propagation of correlations along the separation distance $r$ and provide an adequate representation of the energy cascade, yielding skewness values of the longitudinal velocity derivative and of the temperature derivative equal to $-3/7$ and $-1/5$, respectively.

\bigskip

\section{Background: Kinematics of Continuum Fluids \label{kinematics}}

To investigate certain statistical properties of fully developed turbulence, we first revisit the fundamentals of continuum fluid kinematics following the classical theoretical formulation \cite{Truesdell77}.
These foundational concepts, concerning the various representations of motion, provide the framework necessary for the analysis developed in the present work.
\begin{figure} 
	\centering
	\includegraphics[width=80mm,height=70mm]{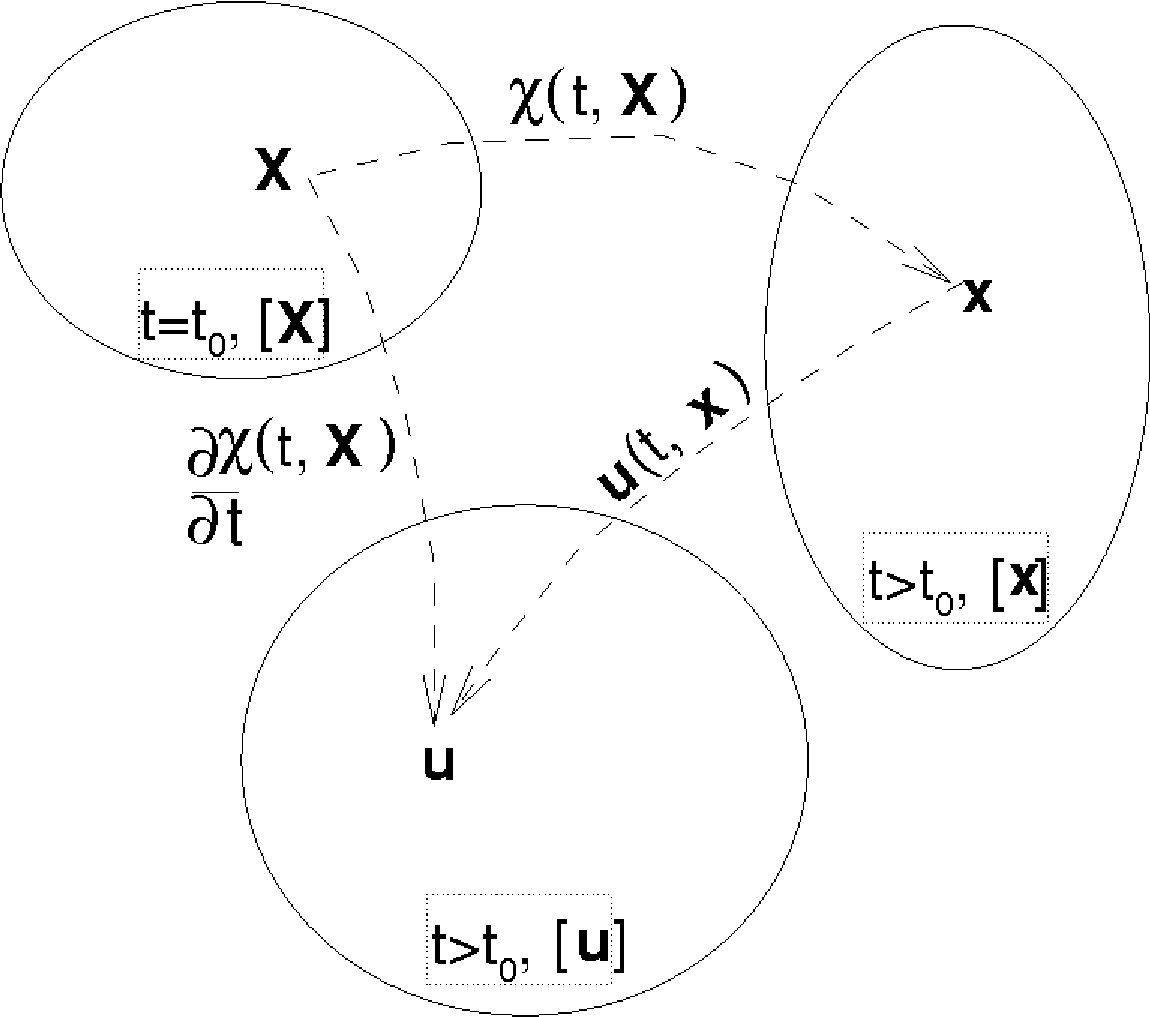}
\caption{Scheme of fluid displacement and velocity field.}
\label{figura_1}
\end{figure}

\begin{figure}
	\centering
	\includegraphics[width=80mm,height=70mm]{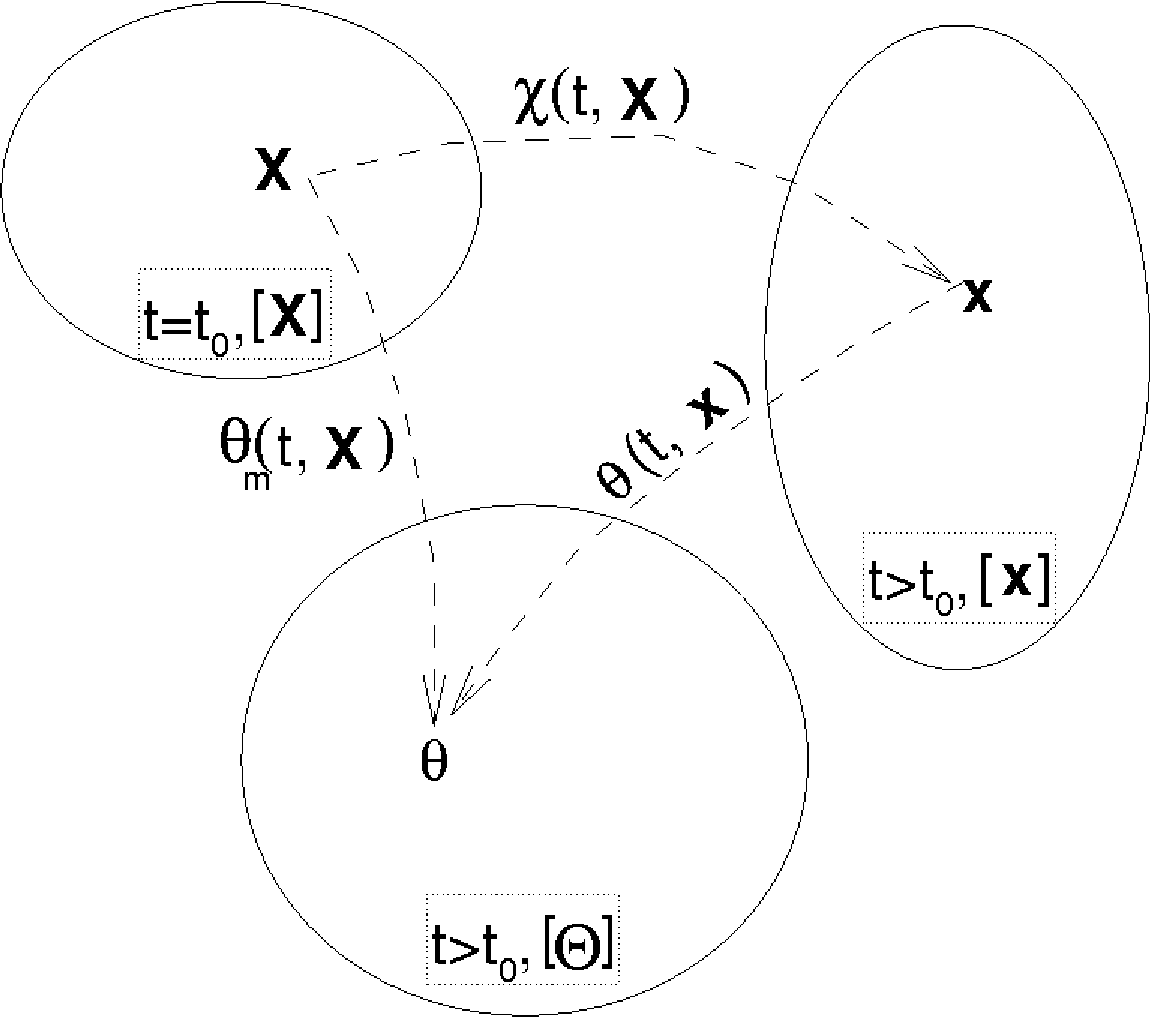}
\caption{Scheme of fluid displacement and temperature field.}
\label{figura_2}
\end{figure}

The following mapping (see the schematics in Figs. \ref{figura_1} and \ref{figura_2}) \cite{Truesdell77}
\bea
\ds {\bfchi}(t,\cdot): {\bf X} \rightarrow {\bf x}(t)
\eea
defines the referential representation of motion, assigning the position at the current time
$t$ $\ne$ $t_0$ of a fluid particle that occupied the referential position $\bf X$ (referential positions) at $t$=$t_0$. 
Accordingly, $\bf X$ also serves as a label uniquely identifying the particle located at $\bf X$ at the reference time $t$=$t_0$.
More generally, the referential configuration denotes the region occupied by the fluid at $t$=$t_0$ or the region it could occupy. Conversely, $\bf X$ can be formally expressed in terms of $\bf x$, through the inverse mapping $\bfchi^{-1}$  \cite{Truesdell77} 
\bea
\begin{array}{l@{\hspace{-0.cm}}l}
\ds {\bf x} = {\bfchi}(t, {\bf X}) \\\\
\ds {\bf X} = {\bfchi}^{-1}(t, {\bf x}) 
\end{array}
\eea
The velocity of $\bf X$ is then defined as 
\bea
\ds \dot{\bfx} \equiv \dot{\bfchi} \equiv \frac{\partial \bfchi}{\partial t} (t, {\bf X})
\label{v material}
\eea
and the temperature of the material element $\bf X$ is written as
\bea
\vartheta_m = \vartheta_m(t, {\bf X})
\label{t material}  
\eea
Equations (\ref{v material})--(\ref{t material}) give Lagrangian or referential representation
of motion, being $\dot{\bfchi}(t, {\bf X})$ and $\vartheta_m(t, {\bf X})$ velocity and temperature variations along the trajectory of $\bf X$. 
Following the Eulerian view point, velocity and temperature fields, ${\bf u}(t, {\bf x})$ and 
$\vartheta(t, {\bf x})$, are defined according to the schemes of Figs. \ref{figura_1} and \ref{figura_2}, by means of $\dot{\bfchi}$, $\vartheta_m$ and the map $\bfchi^{-1}$
\bea
\begin{array}{l@{\hspace{-0.cm}}l}
\ds {\bf u}(t, {\bf x})= \frac{\partial{\bfchi}}{\partial t}\left( t, {\bfchi}^{-1}(t, {\bf x}) \right)
\\\\
\ds \vartheta(t, {\bf x})= \vartheta_m\left(t, {\bfchi}^{-1}(t, {\bf x}) \right)  
\end{array}
\label{fields}
\eea
Such Eulerian fields are defined starting from Eqs. (\ref{v material})--(\ref{t material}), as restrictions of $\dot{\bfchi}(t, {\bf X})$ and $\vartheta_m(t, {\bf X})$ on the motion ${\bfchi}(t, {\bf X})$ \cite{Truesdell77}.
Viceversa, the Lagrangian representation of motion is formulated in terms of Eulerian fields and displacement $\bfchi$.
\bea
\begin{array}{l@{\hspace{-0.cm}}l}
\ds \dot{\bfchi}(t, {\bf X})= {\bf u} \left( t, {\bfchi}(t, {\bf X}) \right), \\\\
\ds \vartheta_m(t, {\bf X})= \vartheta\left(t, {\bfchi}(t, {\bf X}) \right)
\label{t fields}
\end{array}
\eea

\bigskip

\section{Background: Evolution equations, phase space and motion descriptions \label{Background}}

To analyse fluid motion through the Liouville theorem,  
this section first introduces the evolution equations of the fluid state variables and
the associated phase spaces. To this end, the governing equations are formulated on an infinite domain for both the Eulerian and Lagrangian representations, where the motion is described with respect to an assigned inertial frame $\cal R$.

The Eulerian representation describes the dynamics through fluid properties depending on $t$ and ${\bf x}$. In this setting, the mass and momentum equations (Navier--Stokes equations) and the heat equation read
\bea
\begin{array}{l@{\hspace{-0.cm}}l}
\ds \nabla_{\bf x} \cdot {\bf u} = 0, \\\\
\ds \frac{\partial {\bf u}}{\partial t} =
-  \nabla_{\bf x} {\bf u} \ {\bf u} - \frac{\nabla_{\bf x} p}{\rho}  + \nu \nabla_{\bf x}^2 {\bf u},
\end{array}
\label{NS_eq Euler}
\eea
\bea
\begin{array}{l@{\hspace{-0.cm}}l}
\ds \frac{\partial \vartheta }{\partial t} =
-{\bf u} \cdot \nabla_{\bf x} \vartheta + \kappa \nabla_{\bf x}^2 \vartheta,
\end{array}
\label{T_eq Euler}
\eea
where the pressure $p$, eliminated through the continuity equation, is expressed as a functional of 
${\bf u}(t,{\bf x})$:
\bea
\ds  p(t, {\bf x}) =  \frac{\rho}{4 \pi}
\int_{\mathcal V'} \frac{\partial^2 (u_i' u_j')}{\partial x_i' \partial x_j'} \ \frac{d V({\bf x}')}{\vert {\bf x}' - {\bf x} \vert}.
\label{pressure}
\eea
Consequently, the momentum equations take the form of integro--differential equations, where $p$ exerts a nonlocal influence on the dynamics \cite{Tsinober2009}.

Thus, ${\bf u}(t,{\bf x})$ and $\vartheta(t,{\bf x})$ constitute the state variables of the Eulerian description. According to Eq.~(\ref{fields}), these fields are linked to ${\bfchi}(t,{\bf X})$, although the latter does not enter explicitly in Eqs.~(\ref{NS_eq Euler}) and (\ref{T_eq Euler}); hence ${\bfchi}(t,{\bf X})$ is not a state variable in the Eulerian viewpoint, as its evolution is implicitly encoded in Eq.~(\ref{NS_eq Euler}).

Conversely, the Liouville theorem for fluid motion requires adopting the Lagrangian viewpoint, where material properties are evaluated by tracking each fluid particle along its trajectory. In this description, the governing equations are
\bea
\ds \dot{\bfchi} \equiv {\bf u}\left( t, {\bfchi} \right),
\label{chi}
\eea
\bea
\begin{array}{l@{\hspace{-0.cm}}l}
\ds \ddot{\bfchi} \equiv \frac{D{\bf u}}{Dt} \equiv \frac{\partial {\bf u}}{\partial t} + \nabla_{\bf x} {\bf u} \ {\bf u}
= - \frac{\nabla_{\bf x} p}{\rho} + \nu \nabla_{\bf x}^2 {\bf u},
\end{array}
\label{NS_eq Lagrange}
\eea
\bea
\begin{array}{l@{\hspace{-0.cm}}l}
\ds \frac{\partial \vartheta_m}{\partial t} \equiv \frac{D\vartheta}{Dt}
\equiv \frac{\partial \vartheta}{\partial t} + {\bf u} \cdot \nabla_{\bf x} \vartheta
= \kappa \nabla_{\bf x}^2 \vartheta,
\end{array}
\label{T_eq Lagrange}
\eea
where Eq.~(\ref{chi}) governs the evolution of particle position, so that $\bfchi$ is a state variable of the Lagrangian representation, and $p$ is formally given by Eq.~(\ref{pressure}).

Therefore, the state variables in the Lagrangian viewpoint are 
${\bfchi}(t,{\bf \circ})$, ${\bf u}(t,{\bfchi})$, and $\vartheta(t,{\bfchi})$.  
Following Refs.~\cite{deDivitiis_5, deDivitiis_8}, Eqs.~(\ref{NS_eq Lagrange})--(\ref{T_eq Lagrange}) can be recast in the symbolic operator form
\bea
\begin{array}{l@{\hspace{-0.cm}}l}
\ds \dot{\bfchi} = {\bf u}(t,{\bfchi}),
\end{array}
\label{chi bis}
\eea
\bea
\begin{array}{l@{\hspace{-0.cm}}l}
\ds \dot{\bf u} = {\bf N}({\bf u};\nu),
\end{array}
\label{NS_op}
\eea
\bea
\begin{array}{l@{\hspace{-0.cm}}l}
\ds \dot{\vartheta} = {\bf M}(\vartheta;\kappa),
\end{array}
\label{T_op}
\eea
where $\dot{\bf u}$ and $\dot{\vartheta}$ denote the acceleration and the material temperature rate of particles located at ${\bfchi}(t,{\bf \circ})$.  
The operator ${\bf N}$ is quadratic and includes, among other contributions, the integral nonlinear operator that provides $\nabla_{\bf x}p$ as a functional of ${\bf u}$:
\bea
\begin{array}{l@{\hspace{-0.cm}}l}
\ds {\bf N} = {\bf L}\,{\bf u} + \frac{1}{2}\,{\bf C}\,{\bf u}\,{\bf u},
\end{array}
\label{Nw}
\eea
where ${\bf L}$ is the linear operator associated with viscous forces, and ${\bf M}$ depends on $\kappa$ and acts linearly on $\vartheta$.

Equations (\ref{NS_op})--(\ref{T_op}), defined in the infinite--dimensional space of Eulerian fields, do not depend explicitly on ${\bf x}$. To analyse the bifurcations of such equations, we assume that the theory of ordinary differential equations extends to this infinite--dimensional phase space. 
{Specifically, we assume that the phase space of Eqs. (\ref{NS_op})--(\ref{T_op}), though inherently an infinite--dimensional vector space, may be effectively represented by a finite--dimensional manifold. Under this assumption, Eqs. (\ref{NS_op})--(\ref{T_op}) are reduced to the dynamical system framework pioneered by Ruelle and Takens \cite{Ruelle71}. Consequently, the governing equations are analyzed through the lens of classical bifurcation theory for ordinary differential equations; the results presented herein are thus to be considered valid within the formal limits and scope of the Ruelle-–Takens formulation.
From the mathematical perspective, the transition from an infinite--dimensional phase space to a finite--dimensional one can be justified by the convergence of the functional series representing the solution. Provided that the truncation satisfies criteria for uniform convergence (i.e. Weierstrass criterium M-test), the bifurcation analysis performed on the reduced system remains asymptotically consistent with the dynamics of the continuous operator. In this framework, the finite--dimensional space effectively captures the essential analytical structure of the solutions, as the omitted degrees of freedom--residing in the "tail" of the convergent series do not qualitatively alter the stability or the branching evolution of the solutions. 

From the perspective of laminar--turbulent transition, while the Navier--Stokes equations in an infinite domain are inherently infinite-dimensional, the bifurcation analysis can be effectively projected onto a finite--dimensional subspace by invoking the existence of a local center manifold. This is because the onset of instability is driven by a discrete set of critical modes that capture the essential qualitative changes in the flow topology. By assuming that the remaining infinite degrees of freedom exhibit sufficiently strong decay, we can treat the resulting dynamics as being governed by a low--dimensional system of ordinary differential equations, effectively reducing the infinite--dimensional phase space to its most influential degrees of freedom. This reduction effectively compresses the infinite--dimensional phase space into its most influential degrees of freedom without loss of the underlying bifurcation structure.

On the other hand, from the fully developed turbulence point of view, although the Navier--Stokes equations in an infinite domain theoretically possess an infinite number of degrees of freedom, the presence of physical viscosity suggests the existence of a finite--dimensional inertial manifold. In the chaotic regime, the small--scale turbulent fluctuations are nonlinearly slaved to the large-scale dynamics. Therefore, we justify the use of a finite--dimensional phase space as an effective
framework for characterizing the core bifurcation sequence and the transition to chaos, accurately capturing the governing instabilities without the need to resolve the entire dissipative range.}

\bigskip

For what concerns viscosity and thermal diffusivity, $\nu$ and $\kappa$, these are taken to be temperature--independent; therefore, in both representations, the momentum equations are autonomous with respect to the heat equation, whereas the solutions of Eqs.~(\ref{T_eq Euler}) and (\ref{T_eq Lagrange}) depend on Eqs.~(\ref{NS_eq Euler}) and (\ref{NS_eq Lagrange}), respectively.

Finally, although Eq.~(\ref{T_op}) is formulated for temperature, the present analysis applies, without loss of generality, to any passive scalar with diffusive transport.

\bigskip

\section{Lyapunov analysis and bifurcations of Navier--Stokes equations \label{section L1}}

In this section, we present the Lyapunov analysis of Eqs. (\ref{NS_op})–-(\ref{T_op}), together with a detailed investigation of bifurcations in the framework of the relative phase space. Since the passive scalar equation (\ref{T_op}) is linear with respect to $\vartheta$, both transition and turbulence originate from the instabilities and bifurcations of Eq. (\ref{NS_op}), where $\nu^{-1}$ acts as the control parameter. Accordingly, the Lyapunov formulation is here applied to the Navier–-Stokes equations in the relative phase space $\left\lbrace {\bf u} \right\rbrace$. 
Lyapunov exponents and vectors are defined as follows: 
\bea 
\begin{array}{l@{\hspace{-0.cm}}l} 
\ds \dot{\bf u} = {\bf N}({\bf u}; \nu), \\\\ 
\ds \dot{\bf y} = {\bf A}({\bf u}){\bf y}, \ \ \ {\bf y} = {\bf u}^* - {\bf u}, \\\\
\ds \Lambda_{NS} = \frac{\left( {\bf y} \cdot \dot{\bf y} \right)_E}{\left( {\bf y} \cdot {\bf y} \right)_E}, \\\\
\ds \Lambda^{(i)}_{NS} = \left( {\bfeps}^{(i)} \cdot {\bf A}({\bf u}){\bfeps}^{(i)}\right)_E,  \ \ \ 
\ds {\bfeps}^{(i)} = \frac{{\bf y}^{(i)}}{\vert {\bf y}^{(i)} \vert}, \quad i = 1,2,\ldots 
\end{array} 
\label{Lyapunov NS vect & exp} 
\eea 
where ${\bf u}^*$ is the perturbed Eulerian velocity field, $(\circ \cdot \circ )_E$ denotes the inner product defined in $\left\lbrace {\bf u} \right\rbrace$, and the norm $\vert \circ \vert$ is induced by such inner product. The Jacobian $\bf A \equiv \nabla_{\bf u}{\bf N}$ is linear with respect to $\bf u$: 
\bea \begin{array}{l@{\hspace{-0.cm}}l} 
\ds {\bf A} \equiv \nabla_{\bf u}{\bf N} = {\bf L} + {\bf C_s}{\bf u} 
\end{array} 
\label{grad N}
 \eea
 Here, ${\bf C_s}{\bf u}$ denotes the symmetric part of ${\bf C}{\bf u}$, and ${\bf C_s}$ is a constant operator. The vector ${\bf y}$ is the classical Lyapunov vector defined in Eulerian function space, and $\Lambda_{NS}$ is the corresponding Lyapunov exponent. The quantities ${\bfeps}^{(i)}$ and $\Lambda^{(i)}_{NS}$ are, respectively, the canonical Lyapunov unit vectors forming the so-called Lyapunov basis, and the associated canonical Lyapunov exponents. This basis changes its orientation and position in the Eulerian field space as the dynamics evolves. In particular, ${\bfeps}^{(1)}$ identifies the direction of the maximal exponent $\Lambda^{(1)}_{NS}$; ${\bfeps}^{(2)}$, lying in the subspace orthogonal to ${\bfeps}^{(1)}$, provides the second maximal exponent $\Lambda^{(2)}_{NS}$; ${\bfeps}^{(3)}$ gives $\Lambda^{(3)}_{NS}$ in the subspace orthogonal to both ${\bfeps}^{(1)}$ and ${\bfeps}^{(2)}$, and so forth.

Because of fluid dissipation,
\bea 
\ds \sum_{k} \Lambda^{(k)}_{NS} \equiv \mbox{tr}({\bf A}) < 0 
\eea 
In fully developed turbulence, the leading Lyapunov exponents are positive up to a certain index. In particular, $0 < \Lambda^{(1)}_{NS} \lesssim C \Lambda_{L}$, where $\Lambda_{L} \sim \sqrt{\varepsilon/\nu} \sim u/\lambda_T$ represents the mean Lyapunov exponent of the velocity gradient, with $C=O(1)$,
$\varepsilon = \nu \left\langle \partial u_i/\partial x_j \ \partial u_i/\partial x_j \right\rangle_E$ is the energy dissipation rate, $\lambda_T$ is the Taylor microscale, and $\langle \circ \rangle_E$ denotes an Eulerian ensemble average.

Eulerian fields, Lyapunov vectors, and the corresponding exponents exhibit fluctuations due to the bifurcations encountered along the dynamics. To analyse this, we treat bifurcations of Eqs. (\ref{NS_op})–-(\ref{T_op}) following the framework of finite-dimensional ordinary differential equations. According to Ref. \cite{Ruelle71}, we assume that the infinite-dimensional phase space can be treated as a finite-dimensional manifold. Within the validity limits of Ref. \cite{Ruelle71}, this allows the formal application of classical bifurcation theory \cite{Ruelle71, Eckmann81, Guckenheimer90} to Eqs. (\ref{NS_op})–-(\ref{T_op}).

Such bifurcations occur at points of $\left\lbrace {\bf u} \right\rbrace$ where the Jacobian ${\nabla_{\bf u}{\bf N}}$ has at least one eigenvalue with zero real part (NS-bifurcations), i.e. 
\bea
 \ds \Sigma_u : \ {\cal D}_{NS} \equiv \det\left( \nabla_{\bf u}{\bf N} \right) = 0.
 \label{det NS} 
\eea
 Expression (\ref{det NS}) represents a secular condition defining a hypersurface $\Sigma_u \subset \left\lbrace {\bf u} \right\rbrace$. As the underlying phase space is infinite-dimensional with a global norm, $\Sigma_u$ is expected to be smooth. Along the dynamics, the repeated vanishing of ${\cal D}_{NS}$ induces fluctuations in ${\bf u}$, $\Lambda_{NS}$, and ${\bf y}$, whose rapidity is governed by the rate at which Navier--Stokes bifurcations occur. This is the mean rate of crossing of the Jacobian--degeneracy manifold, defined as 
\bea
\begin{array}{l@{\hspace{-0.cm}}l}
\ds S_{NS} = \lim_{T \rightarrow \infty} \frac{1}{T}\int_0^T \delta({\cal D}_{NS}) \left\vert \frac{d{\cal D}_{NS}}{dt} \right\vert dt 
\end{array} 
\label{NS rate} 
\eea
and corresponds to the average frequency with which trajectories of Eq. (\ref{NS_op}) intersect $\Sigma_u$. Here, 
\bea
 \begin{array}{l@{\hspace{-0.cm}}l} 
\ds \frac{d{\cal D}_{NS}}{dt} = {\cal D}_{NS} \ \mbox{tr} \left( {\bf A}^{-1}\frac{d{\bf A}}{dt} \right), \\\\
\ds \frac{d{\bf A}}{dt} = (\nabla_{\bf u}{\bf A}) \dot{\bf u} = {\bf C_s} \dot{\bf u} 
\end{array}
\label{app}
\eea
 Since ${d{\bf A}}/{dt}$ depends linearly on fluid-particle acceleration and ${\bf C_s}$ is constant, Eqs. (\ref{app})–-(\ref{NS rate}) suggest the following estimate for the order of magnitude of $S_{NS}$: 
\bea
 \ds S_{NS} \sim \frac{({\bf u}\cdot \dot{\bf u})_E}{({\bf u}\cdot{\bf u})_E}
 \label{ht}
 \eea
 Equation (\ref{ht}) gives an estimate of $S_{NS}$ averaged over the entire infinite-dimensional phase space. In homogeneous turbulence, $S_{NS}$ reduces to the local rate of $u$, namely 
\bea 
\begin{array}{l@{\hspace{-0.cm}}l} 
\ds S_{NS} \backsim \left\vert \frac{d}{dt}\ln u \right\vert,
 \end{array}
 \label{hit}
 \eea
 where $u$ is the local velocity standard deviation defined according to \cite{Karman38}. Using the evolution equation of $u$ in homogeneous isotropic turbulence \cite{Karman38}, $S_{NS}$ may be expressed in terms of viscosity and the Taylor scale $\lambda_T$: 
\bea
 \begin{array}{l@{\hspace{-0.cm}}l} 
\ds S_{NS} \backsim \frac{\nu}{\lambda_T^2} = \frac{u}{\lambda_T}\frac{1}{R_T} \backsim \frac{\Lambda^{(1)}_{NS}}{R_T}, \\\\
\ds R_T = \frac{u\lambda_T}{\nu} 
\end{array}
 \label{P S_NS}
 \eea
 where $R_T$ is the Taylor-scale Reynolds number. On the other hand, $\Lambda^{(1)}_{NS}$ depends on Reynolds number according to \cite{Ruelle1979}: 
\bea
 \begin{array}{l@{\hspace{-0.cm}}l} 
\ds \Lambda^{(1)}_{NS} \backsim \frac{u}{L} Re^n, \\\\
\ds n = \frac{1-m}{1+m}, \\\\
\ds Re = \frac{uL}{\nu}, \qquad \frac{L}{\lambda_T} \sim R_T, 
\end{array} 
\eea 
where $L$ is the integral scale of longitudinal correlation function, and $m$ is the H\"older exponent of the velocity increments, 
$m \sim \ln(|\Delta u_2|/|\Delta u_1|)/\ln(r_2/r_1)$. According to Kolmogorov \cite{Kolmogorov41}, $m=1/3$, hence $n=1/2$.

From Eq. (\ref{P S_NS}), the bifurcation rate of the Navier–-Stokes equations is significantly smaller than the growth rate associated with the maximal Lyapunov exponent. Therefore, the fluctuations induced by such bifurcations on $\Lambda^{(1)}_{NS}$, on the other exponents, on ${\bf y}$, and on Eulerian fields evolve at rates much slower than the exponential separation associated with $\Lambda^{(1)}_{NS}$. Moreover, as $R_T \rightarrow \infty$, both $\Lambda^{(1)}_{NS}$ and $S_{NS}$ diverge, but with different growth rates.

Navier–-Stokes bifurcations also generate a doubling of velocity and temperature fields, meaning that all properties associated with these fields including their characteristic scales $\ell_q$ and times $\tau_q$, $q = 1,2,\ldots$—are duplicated \cite{deDivitiis_8}. Here, $q$ denotes the number of NS bifurcations encountered as $\nu\rightarrow 0$ acts as the control parameter of Eq. (\ref{NS_op}). After $q$ bifurcations, the velocity field takes the form: 
\bea
 \begin{array}{l@{\hspace{-0.cm}}l}
 \ds {\bf u}(t,{\bf x}) = {\bf u} \left( \frac{t}{\tau_1},\frac{t}{\tau_2},\ldots,\frac{t}{\tau_q}; \ \frac{\bf x}{\ell_1},\frac{\bf x}{\ell_2},\ldots,\frac{\bf x}{\ell_q} \right), 
\end{array}
 \label{u l}
 \eea
 where, according to \cite{Ruelle71, Feigenbaum78, Pomeau80, Eckmann81}, turbulence begins when $q\gtrsim 3,4$, and subsequently $q$ diverges while $\ell_k$ and $\tau_k$ become essentially continuously distributed.

In fully developed turbulence, the smallest of these scales, $\ell_{\min} \equiv \min{\ell_k} \equiv \ell_m$, corresponds to the Kolmogorov scale, determined by the local balance between inertial and viscous forces:
 \bea
 \begin{array}{l@{\hspace{-0.cm}}l}
 \ds \frac{u_k,\ell_{\min}}{\nu} \approx 1, \ \
\ds \ell_{\min} = u_k,\tau_m, \ \ 
\ds \lambda_T = u,\tau_m. 
\end{array}
 \label{small scale}
 \eea
 Eliminating $\tau_m$ from Eqs. (\ref{small scale}) yields 
\bea
 \begin{array}{l@{\hspace{-0.cm}}l}
 \ds \frac{\lambda_T}{\ell_{\min}} \approx \frac{u}{u_k} \approx \sqrt{R_T} 
\end{array} 
\label{Taylor kolmogorov}
 \eea 
showing that $\ell_{\min}$ and $u_k$ represent, respectively, the Kolmogorov length scale and the associated velocity scale.

The trajectory of an arbitrary fluid particle ${\bf X}$, obtained from Eq. (\ref{u l}), 
\bea
 \begin{array}{l@{\hspace{-0.cm}}l}
 \ds \dot{\bfchi}(t,{\bf X}) = {\bf u} \left( \frac{t}{\tau_1},\frac{t}{\tau_2},\ldots,\frac{t}{\tau_q}; \ \frac{\bfchi(t,{\bf X})}{\ell_1},\frac{\bfchi(t,{\bf X})}{\ell_2},\ldots,\frac{\bfchi(t,{\bf X})}{\ell_q} \right), 
\end{array}
 \label{traj}
 \eea
 is therefore expected to be significantly more rapid and irregular than the Eulerian field fluctuations thanks to the nested functional structure of the velocity field. This induces sharp, rapid variations in Lagrangian quantities, especially in the velocity gradient along particle trajectories. These effects arise solely from the Navier–-Stokes bifurcations acting on the Lagrangian velocity field.

We conclude by noting that Eq. (\ref{P S_NS}) and its implications apply to fully developed homogeneous isotropic turbulence.

\bigskip

\section{Lyapunov analysis and bifurcations in the Lagrangian set \label{section L2}}

In this section we develop a finite--scale Lyapunov analysis of fluid--particle trajectories and introduce a precise characterization of bifurcation phenomena in the Lagrangian description of motion.
Finite--scale Lyapunov exponents and vectors are defined through the following dynamical system:
\bea
\begin{array}{l@{\hspace{-0.cm}}l} 
\ds \dot{\bfchi} = {\bf u}(t, {\bfchi}), \\\\
\ds {\bf x} = {\bfchi}(t, {\bf X}), \\\\ 
\ds \dot{\bfxi} = {\bf u}(t, {\bfchi}+{\bfxi}) - {\bf u}(t, {\bfchi}), \\\\ 
\ds \Lambda_{L} = \frac{{\bfxi} \cdot \dot{\bfxi}}{{\bfxi}\cdot{\bfxi}}, \\\\
 \ds \Lambda^{(i)}_{L} = \frac{{\bf e}^{(i)}}{\vert\bfxi\vert} \cdot \left( {\bf u}(t, {\bfchi}+\vert{\bfxi}\vert {\bf e}^{(i)}) - {\bf u}(t, {\bfchi}) \right), \ \ i = 1, 2, 3,
\end{array} 
\label{Lyapunov v vect & exp} 
\eea
where $\bfxi$ denotes the finite--scale Lyapunov separation vector in physical space and $\Lambda_L$ the associated instantaneous growth rate.
The vectors ${\bf e}^{(i)}$ and the corresponding exponents $\Lambda^{(i)}_L$, $i=1,2,3$, define the finite--scale Lyapunov basis.
This basis is advected along the material trajectory ${\bf X}$ and continuously reoriented so that ${\bf e}^{(1)}$ identifies the direction of maximal stretching, ${\bf e}^{(2)}$ lies in the orthogonal plane and yields the second exponent, and ${\bf e}^{(3)}$ completes the orthonormal triad.
The classical Lyapunov exponents are recovered in the limit $\vert\bfxi\vert \rightarrow 0$.

Owing to the strong spatio--temporal intermittency of the velocity gradient tensor $\nabla_{\bf x}{\bf u}(t,{\bf x})$, the quantities $\bfxi$, ${\bf e}^{(i)}$ and $\Lambda^{(i)}_L$ undergo intense fluctuations along particle trajectories.
To characterize their origin, we now introduce the notion of \emph{Lagrangian bifurcations}.

A Lagrangian bifurcation occurs whenever the velocity gradient tensor admits at least one eigenvalue with vanishing real part, equivalently when its determinant vanishes:
\bea
\Sigma_L:\ {\cal D}_L \equiv \det \left( \nabla_{\bf x} {\bf u}(t, {\bf x}) \right) = 0.
\label{det Grad v}
\eea
Equation~(\ref{det Grad v}) defines a surface $\Sigma_L \subset \mathbb{R}^3$ representing the locus of local Lagrangian bifurcations.
Due to the nested functional structure of the velocity field and the irregularity of $\nabla_{\bf x}{\bf u}$, $\Sigma_L$ is an unsteady, highly non--smooth surface, characterized at each instant by folds, thin layers and sharp gradient variations.

The \emph{Lagrangian bifurcation rate} $S_L$ is defined as the frequency with which a particle trajectory intersects $\Sigma_L$:
\bea
\begin{array}{l@{\hspace{-0.cm}}l}
\ds S_L = \lim_{T\rightarrow\infty}\frac{1}{T}\int_0^T \delta({\cal D}_L) \left| \frac{D{\cal D}_L}{Dt} \right| \ dt,
\end{array}
\label{kin rate}
\eea
where the material derivative reads
\bea
\frac{D\circ}{Dt} = \frac{\partial\circ}{\partial t}
+
\nabla_{\bf x}\circ \cdot {\bf u}.
\eea
Furthermore,
\bea
\frac{D {\cal D}_L}{Dt}
=
{\cal D}_L\,
\mbox{tr}
\left(
\left( \nabla_{\bf x}{\bf u} \right)^{-1}
\frac{D (\nabla_{\bf x}{\bf u})}{Dt}
\right).
\eea
Using the multiscale structure of turbulence and the extreme geometrical complexity of $\Sigma_L$, one expects
\bea
\left| \frac{\partial {\cal D}_L}{\partial t} \right|
\ll
\left| \nabla_{\bf x} {\cal D}_L \cdot {\bf u} \right|,
\label{DK}
\eea
which leads to the estimate
\bea
\ds S_L \backsim \frac{u}{\ell_{\min}}
\sim
\frac{u}{\lambda_T}\sqrt{R_T}
\sim
\Lambda_L^{(1)}\sqrt{R_T}.
\label{rate S_L}
\eea

Equation~(\ref{rate S_L}) shows that Lagrangian bifurcations occur at rates much faster than the exponential separation governed by $\Lambda_L^{(1)}$.
As a consequence, fluctuations of the finite--scale Lyapunov exponents take place on time scales far shorter than those associated with trajectory divergence.
Both $S_L$ and $\Lambda_L^{(1)}$ diverge as $R_T \rightarrow \infty$, but with markedly different scaling behavior.

\section{Remark: Lagrangian versus Eulerian regimes \label{section L3}}

The previous analysis allows us to establish a sharp distinction between Lagrangian and Eulerian fluctuation regimes in fully developed turbulence.
Specifically,
\bea
\begin{array}{l@{\hspace{-0.cm}}l}
\underbrace{\ds S_L >> 
\sup\left\lbrace \Lambda_L \right\rbrace >>
\left\langle \Lambda_L \right\rangle_L }\gtrsim 
\underbrace{\sup\left\lbrace \Lambda_E \right\rbrace >>
\left\langle \Lambda_E \right\rangle_E >>
S_E} \\\
 \mbox{Lagrangian \ parameters} \ \ \ \  \  \ \ \ \mbox{Eulerian \ parameters}
\end{array}
\label{S>>>>S}
\eea
where $\langle\cdot\rangle_L$ and $\langle\cdot\rangle_E$ denote averages over the Lagrangian and Eulerian sets, respectively.

Combining Eq.~(\ref{rate S_L}) with Eq.~(\ref{P S_NS}) yields
\bea
\ds \frac{S_L}{S_{NS}} \backsim R_T^{3/2}.
\label{SK/SNS}
\eea
Since homogeneous isotropic turbulence requires $R_T \gtrsim 10$ \cite{Batchelor53}, one finds
\bea
\begin{array}{l@{\hspace{-0.cm}}l}
\ds \frac{S_L}{S_{NS}} \gg 10, \qquad
\inf_{R_T}\left\{\frac{S_L}{S_{NS}}\right\} \backsim 40.
\end{array}
\label{rates}
\eea

Equations~(\ref{S>>>>S})--(\ref{rates}) show that fluctuations along particle trajectories are substantially more rapid than those of the Eulerian field.
Two qualitatively distinct regimes therefore emerge:
(i) an Eulerian regime dominated by stretching, in which Lyapunov exponents fluctuate on time scales slower than perturbation growth;
(ii) a Lagrangian regime dominated by very rapid folding mechanism, in which Lyapunov exponents fluctuate much faster than trajectory separation.

This disparity reflects the fact that Lagrangian dynamics directly samples intense local gradients of the velocity field, whereas the Navier--Stokes evolves in a high--dimensional phase space constrained by a global norm.
Similarly, the surface $\Sigma_L \subset \mathbb{R}^3$ is far more tortuous than its Eulerian counterpart $\Sigma_{NS}$, leading to a much higher frequency of local bifurcations along Lagrangian trajectories.

\bigskip

\section{Liouville spectral analysis and local bifurcation rate \label{section L4}}

To study the correlation between Eulerian and Lagrangian fields, we first present a specific Liouville analysis of a given dynamical system together with a detailed investigation of bifurcations in the framework of the relative phase space. 
To establish a quantitative link between bifurcation dynamics and the decay of Eulerian--Lagrangian correlations, we consider a deterministic dynamical system
\bea
\dot{\mathbf{z}} = {\mathbf{R}}(\mathbf{z}),
\eea
where $F(t,\mathbf{z})$ denotes the phase--space probability density function (PDF), whose evolution is governed by the Liouville equation
\bea
\ds \frac{\partial F}{\partial t} = - \nabla F \cdot \mathbf{R} - F \, \nabla \cdot \mathbf{R}.
\label{eq:liouville_pdf}
\label{liouville 2}
\eea
The connection between bifurcation dynamics and correlation decay originates from the first term on the right--hand side of Eq.~(\ref{liouville 2}),
which governs the advection of probability gradients along the flow,
while the second term accounts solely for the contraction or expansion of infinitesimal phase--space volumes.

In a strongly chaotic regime, the local bifurcation rate admits a purely kinematic estimate.
Specifically, one may define
\bea
\begin{array}{l@{\hspace{-0.cm}}l}
\ds S_{\bf Z} = \frac{1}{\Delta t_{\bf Z}}= \frac{1}{2}
  \left| \frac{\nabla D \cdot \nabla {\bf R} {\bf R} + {\bf R \cdot {\bf H} {\bf R}}}{\nabla D \cdot {\bf R} +\partial_t D}\right|_{\bf Z}, \\\\
\ds \Sigma_D: D= \det\left(\nabla {\bf R} \right) = 0, \\\\
\ds \nabla \circ \equiv \left( \frac{\partial \circ}{\partial z_1}, \frac{\partial \circ}{\partial z_2}, ... \right), \ \ {\bf H} = \nabla \nabla D = \left( \left( \frac{\partial^2 D }{\partial z_i \partial z_j}\right) \right),
\end{array}
\label{loc bif rate}
\eea
where $S_{\bf Z}$, the local bifurcation rate, is the inverse of the time interval $\Delta t_{\bf Z}$ between two consecutive intersections, at times $\bar t$ and $\bar t+\Delta t_{\bf Z}$, of a phase--space trajectory with the hypersurface $\Sigma_D$.
The subscript ${\bf Z}$ denotes one of the points ${\bf Z}_k \in \Sigma_D$ at which $S_{\bf Z}$ attains local maxima whose statistical distribution is stationary along ergodic trajectories.
This construction yields a discrete set $\{S_k\}$, whose ensemble average is proportional to the mean bifurcation rate.

Equation~(\ref{eq:liouville_pdf}) can be written in operator form as
\begin{equation}
\ds \frac{\partial F}{\partial t} = \mathcal{L} F,
\qquad
\mathcal{L}
= - \nabla \cdot \mathbf{R}
- \mathbf{R}\cdot\nabla , \qquad
\mathcal{L}^\dagger
= \nabla \cdot \mathbf{R}(.),
\end{equation}
where $\mathcal{L}$ and $\mathcal{L}^\dagger$ denote the Liouville and Koopman operators, respectively.
Let $\phi_k(\mathbf{z})$ and $\psi_k(\mathbf{z})$ be the associated eigenfunctions,
\begin{equation}
\mathcal{L}\phi_k = \lambda_k \phi_k, \qquad
\mathcal{L}^\dagger \psi_k = \lambda_k \psi_k,
\label{eq:adjoint_eigen}
\end{equation}
satisfying $\langle \psi_k , \phi_j \rangle = \delta_{k j}$.
Expanding the PDF as
\begin{equation}
F(t,\mathbf{z})
= \sum_k a_k \phi_k(\mathbf{z}) e^{\lambda_k t}, \ \ \ a_k=\left\langle F, \psi_k \right\rangle 
\end{equation}
shows that the real parts of $\lambda_k$ determine the decay rates of phase--space correlations, where
\bea
\ds {\Re(\lambda_0)}=0> {\Re(\lambda_1)}>{\Re(\lambda_2)}>...
\eea
In this regime, eigenfunctions and eigenvalues, $\phi_k$ and $\lambda_k$ encode the statistical imprint of bifurcation dynamics. Their generic structure reflects the local bifurcation intensities $S_k$.
Substitution of Eq.~(\ref{eq:adjoint_eigen}) yields
\begin{equation}
\lambda_k
= - \mathbf{R}\cdot\nabla \ln \phi_k
- \nabla\cdot\mathbf{R},
\label{eq:lambda_local}
\label{eq:expanded_adjoint}
\end{equation}
which, upon ergodic averaging, gives
\begin{equation}
\Re(\lambda_k)
=
- \left\langle
\mathbf{R} \cdot
\nabla \ln |\phi_k|
\right\rangle
- \left\langle
\nabla\cdot\mathbf{R}
\right\rangle.
\label{eq:lambda_split}
\end{equation}
To establish a connection among $\lambda_k$, $S_k$, and the Lyapunov exponents, observe that, in strongly chaotic regime, due to bifurcations, $\ln \left| \phi_k({\bf z})\right|$, $k = 1, 2...$ are monotonic rising functions of time along an arbitrary phase trajectory.
This is because, due to the combined effect of positive Lyapunov exponents and bifurcations, a flux tube containing phase trajectories stretches and folds continuously in such a way that the same trajectories intersect very frequently a given volume of phase space providing increasingly increasing values of $\left| \phi_k({\bf z})\right|$. Then, we consider two phase--space points, ${\bf z}_1$ and ${\bf z}_2$, with ${t_2>t_1}$, located along a phase trajectory immediately before consecutive bifurcations occur. The travel time required to move from ${\bf z}_1$ to ${\bf z}_2$, i.e. $t_2-t_1$, is therefore the inverse of $S_k$. Accordingly, in the presence of sequential bifurcations, the following inequality holds 
\bea 
\ds \ln \left| \frac{\phi_k({\bf z}_2)}{\phi_k({\bf z}_1)}\right|  =O(1)>0 . 
\eea
Consequently, the second term in Eq.(\ref{eq:lambda_split}) can be rewritten in terms of $S_k$ together with an appropriate contribution of the Lyapunov exponents. In particular, the first term in Eq.(\ref{eq:lambda_split}) incorporates a mode-–dependent mean bifurcation rate, 
\begin{equation} 
\begin{array}{l@{\hspace{-0.cm}}l}
 \ds \left\langle \mathbf{R} \cdot \nabla \ln |\phi_k| \right\rangle \equiv C_k S_k- \sum_{\Lambda_i<0}\left\langle \Lambda_i\right\rangle , \\\\
 \ds  C_k = \ln \left| \frac{\phi_k({\bf z}_2)}{\phi_k({\bf z}_1)}\right|  =O(1),
 \end{array}
\end{equation}
which quantifies the generation of fine--scale structure through repeated splitting and folding of probability filaments, where $C_k$ provides the amplification of $\phi_k$ along the segment of trajectory separating the two bifurcations, and the contribution to the infinitesimal phase--space volumes contraction. The second contribution,
\begin{equation}
\sigma_k
=
\left\langle
\nabla\cdot\mathbf{R}
\right\rangle
=
\sum_{\Lambda_i < 0} \left\langle \Lambda_i\right\rangle 
+
\sum_{\Lambda_i \ge 0} \left\langle \Lambda_i\right\rangle  \equiv \sigma^-_k + \sigma^+_k,
\end{equation}
accounts for net phase--space contraction or expansion and is directly related to the Lyapunov spectrum.
Accordingly,
\begin{equation}
\Re(\lambda_k)
=
- \left( C_k S_k + \sum_{\Lambda_i \ge 0} \left\langle \Lambda_i \right\rangle \right),
\label{decomposition}
\end{equation}
which highlights the complementary roles of bifurcation--induced longitudinal complexity and Lyapunov--driven transverse deformation in controlling correlation decay.

The decomposition (\ref{decomposition}) provides a direct interpretation of the spectral gap of the Liouville operator,
defined as
\begin{equation}
\Delta_{\mathcal{L}}
\equiv
\min_{k\neq 0} \left| \Re(\lambda_k) \right|.
\end{equation}
The gap reflects the competition between the bifurcation rate $S_k$ and the transverse deformation rate $\sigma_k$.
A finite gap therefore requires persistent generation of gradients in the PDF eigenmodes,
balanced by phase--space volume effects.
Rapid correlation decay occurs when bifurcation--induced longitudinal complexity dominates over purely Lyapunov--driven mechanisms.

This interpretation connects the spectral gap to the classical stretching--folding picture:
$\sigma_k$ measures exponential separation or contraction of trajectories,
whereas $S_k$ quantifies repeated splitting and folding of probability filaments.
The spectral gap thus emerges as a statistical measure of the efficiency of the stretching--folding mechanism acting on the invariant density.

Within this framework, the Liouville eigenvalues play the role of Ruelle--Pollicott resonances \cite{Pollicott85, Ruelle86, Ruelle87}.
Unlike the standard formulation, the present approach explicitly resolves the real parts of the resonances into a bifurcation--controlled contribution $S_k$
and a Lyapunov--controlled contribution $\sigma_k$,
clarifying the distinct dynamical origins of correlation decay and providing a direct bridge between  bifurcations and global statistical relaxation.

Finally, we recall the evolution equations for the Eulerian velocity field and for the Lagrangian displacement:
\bea
\dot{\bf u} = {\bf N}({\bf u};\nu),
\eea
\bea
\dot{\bfchi}(t,{\bf X})
=
{\bf u}
\left(
\frac{t}{\tau_1},\ldots,\frac{t}{\tau_q};
\frac{\bfchi}{\ell_1},\ldots,\frac{\bfchi}{\ell_q}
\right).
\label{u 2}
\eea
The Navier--Stokes operator ${\bf N}$ acts in an infinite--dimensional phase space endowed with a global norm, implying that the associated bifurcation surface $\Sigma_u$ is relatively smooth.
By contrast, the nested multiscale structure in Eq.~(\ref{u 2}) renders $\Sigma_L$ highly irregular and strongly folded, producing extremely large local bifurcation rates according to Eq.~(\ref{loc bif rate}).
This explains the dominant role of Lagrangian bifurcations in shaping the Liouville spectrum and the associated Ruelle--Pollicott resonances.

\bigskip

\section{Liouville Analysis of Eulerian--Lagrangian Correlation Decay}

The purpose of the present section is to demonstrate, by means of Liouville theorem, that in fully developed turbulence the Lagrangian and Eulerian fields tend to become statistically uncorrelated, independently of the initial conditions.
To this end, let us consider the trajectory of a single fluid particle $\bf X$.
According to the results discussed in the previous sections, the fluctuations of
${\bfchi}(t, {\bf X})$ are much faster than those of the Eulerian velocity field
${\bf u}(t, {\bf x})$ (see Eqs. (\ref{S>>>>S}) and (\ref{rates})). In particular, during the motion
of $\bf X$, the time interval between two consecutive NS--bifurcations contains a
statistically significant number of Lagrangian bifurcations, especially when
$R_T$ is sufficiently large. Consequently, the variations of
${\bf u}(t, {\bf x})$ are expected to be essentially irrelevant for the statistics
of the fluctuations of $\bfchi$, whose distribution does not depend on the specific
realization of ${\bf u}(t, {\bf x})$.

On the other hand, the distribution of the fluid state variables,
$F$ = $F(t, {\bfchi}, {\bf u}, \vartheta)$, is the joint PDF of the Lagrangian and Eulerian fields.
This PDF satisfies the Liouville equation associated with Eqs. (\ref{chi bis}), (\ref{NS_op}) and (\ref{T_op})
\bea
\ds \frac{\partial F}{\partial t}+
\nabla_{\bf x} F \cdot {\bf u}(t, {\bfchi})+
\nabla_{\bf u} \cdot (F {\bf N})+
\nabla_{\vartheta} \cdot (F {\bf M}) =0 
\label{Liouville}
\eea
where $\nabla_{\bf u}$ and $\nabla_{\vartheta}$ denote functional derivatives, defined in the
infinite-dimensional spaces of functions of the Eulerian fields ${\bf u}$ and $\vartheta$,
respectively.
In operator form, the Liouville equation can be written as
\bea
\ds \frac{\partial F}{\partial t}=\mathcal{L} F
\eea
where the linear Liouville operator $\mathcal{L}$ can be decomposed, in accordance
with Eq. (\ref{Liouville}), as the sum of two operators,
\bea
\ds \mathcal{L}= \mathcal{L}_L + \mathcal{L}_E .
\eea
Specifically,
\bea
\begin{array}{l@{\hspace{-0.cm}}l}
\ds \mathcal{L}_L \circ = -\nabla_{\bf x} \circ \cdot {\bf u}(t, {\bfchi}), \\\\
\ds \mathcal{L}_E \circ = -\nabla_{\bf u} \cdot ( \circ {\bf N} )-
 \nabla_{\bf \vartheta} \cdot ( \circ {\bf M}),
\end{array}
\eea
where both operators $\mathcal{L}_L$ and $\mathcal{L}_E$ depend on the Eulerian velocity field
and are therefore not independent in the general case.
Accordingly, the formal solution of the Liouville equation in terms of the initial
conditions reads
\bea 
\begin{array}{l@{\hspace{-0.cm}}l}
\ds F(t, {\bfchi}, {\bf u}, \vartheta) = \mbox{e}^{t \mathcal{L}} F(0, {\bfchi}, {\bf u}, \vartheta)
= \mbox{e}^{t \mathcal{L}_L}  \mbox{e}^{t \mathcal{L}_E} F(0, {\bfchi}, {\bf u}, \vartheta),
\end{array}
\label{form solut}
\eea
where
\bea 
\begin{array}{l@{\hspace{-0.cm}}l}
\ds \mbox{e}^{t \mathcal{A}} = \Sum_{k=0}^{\infty}  \frac{\mathcal{A}^k}{k!} t^k, \ \
\mathcal{A} = \mathcal{L}, \mathcal{L}_L, \mathcal{L}_E. 
\end{array}
\label{exp oper}
\eea
According to Eq. (\ref{form solut}), the operators
$\mathcal{L}_E$ and $\mathcal{L}_L$ generally act on both Eulerian and Lagrangian variables.
For this reason, Eq. (\ref{form solut}) is not, in general, particularly effective
for characterizing the structure of $F$ at later times in generic fluid-dynamical problems.
However, as established in the previous sections, in fully developed turbulence
the fluctuations of ${\bfchi}(t, {\bf X})$ occur on time scales much shorter than those
associated with ${\bf u}(t, {\bf x})$, due to the corresponding bifurcation rates.
As a consequence, these Lagrangian fluctuations are not affected by the specific
realization of ${\bf u}(t, {\bf x})$.
In this regime, the Lagrangian operator $\mathcal{L}_L$ can therefore be expressed
in terms of an arbitrary, representative realization of the Eulerian velocity field,
denoted by $\hat{\bf u}$, during the turbulent flow,
\bea  
\ds \mathcal{L}_L \circ = -\nabla_{\bf x} \circ \cdot {\bf u}(t, {\bfchi})\equiv 
-\nabla_{\bf x} \circ \cdot \hat{\bf u}(t, {\bfchi}) .
\eea
This crucial observation allows one to conclude that, in fully developed turbulence,
$\mathcal{L}_L$ acts solely on the Lagrangian field $\bfchi$, whereas $\mathcal{L}_E$
operates exclusively on the Eulerian fields.

The PDFs of the Lagrangian and Eulerian fields are then obtained, by definition, as
\bea
\begin{array}{l@{\hspace{-0.cm}}l}
\ds F_E(t, {\bf u}, \vartheta) =\int_{\chi} F \ d {\bfchi}, \\\\
\ds F_L(t, {\bfchi}) = \int_{ {\bf u} \times {\vartheta}} F \ d{\bf u} \ d{\vartheta},
\end{array}
\eea
where $d {\bfchi}$ and $d{\bf u}\, d{\vartheta}$ denote the elemental volumes in the
Lagrangian and Eulerian function spaces, respectively.
The joint PDF can therefore be decomposed as
\bea 
\ds F(t, {\bfchi}, {\bf u}, \vartheta) = F_E(t, {\bf u}, \vartheta) F_L(t, {\bfchi}) + 
\zeta(t, {\bfchi}, {\bf u}, \vartheta),
\eea
where the correlation function $\zeta$ satisfies
\bea
\begin{array}{l@{\hspace{-0.cm}}l}
\ds \int_{\chi} \zeta \ d {\bfchi} =0, \\\\ 
\ds \int_{ {\bf u} \times {\vartheta}} \zeta \ d{\bf u} \ d{\vartheta} =0,
\end{array}
\ \ \ \ \ \ \forall t \ge 0 .
\eea
Accordingly, $\zeta$ obeys the following Liouville equation:
\bea 
\ds \frac{\partial \zeta}{\partial t}- \mathcal{L} \zeta=
\mathcal{L} \left( {F_E}{F_L}\right)  - 
\frac{\partial}{\partial t} \left(F_E F_L\right)
\label{zeta 1}  
\eea
Using the properties of the exponential operator (\ref{exp oper}) and of $\mathcal{L}$,
Eq. (\ref{zeta 1}) can be rewritten as
\bea 
\ds \frac{\partial} {\partial t} \left(\mbox{e}^{-t \mathcal{L}} \zeta \right) =
\mbox{e}^{-t \mathcal{L}} \left( \mathcal{L} \left( {F_E}{F_L}\right)  - 
\frac{\partial}{\partial t} \left(F_E F_L\right)\right),
\label{zeta 2}  
\eea
whose formal solution is
\bea
\begin{array}{l@{\hspace{-0.cm}}l}
\ds  \zeta_t
= 
\mbox{e}^{t \mathcal{L}} \zeta_0
 + \hspace{-1 mm}\int_0^t \hspace{-1 mm} 
\mbox{e}^{(t-\tau) \mathcal{L}}
\left(\mathcal{L} \left( {F_E}{F_L}\right)  - 
\frac{\partial}{\partial \tau} \left(F_E F_L\right)\right)_\tau d \tau ,
\end{array}
\label{zeta_t}
\eea
where the subscripts $0$, $t$, and $\tau$ refer to the corresponding time instants.

To show that $\zeta$ decays to zero on time scales much shorter than the Lyapunov
time $1/\Lambda_L$, independently of the initial conditions, the functions $\zeta_0$,
$F_E$, $F_L$, and their time derivatives are expanded in series of Liouville--Koopman
biorthogonal eigenfunctions of the operators $\mathcal{L}_L$ and $\mathcal{L}_E$,
\bea
\begin{array}{l@{\hspace{-0.cm}}l}
\ds \zeta(0, {\bfchi}, {\bf u}, \vartheta) = \sum_h \sum_k Z_{h k}
\Phi_{L h}({\bfchi}) \Phi_{E k}({\bf u}, \vartheta), \\\\
\ds F_E(\tau, {\bf u}, \vartheta) = \sum_h E_{h} \Phi_{E h}({\bf u}, \vartheta), \ \ 
E_h=\left\langle F_E  \Psi_{E h} \right\rangle_E ,\\\\
\ds \frac{\partial F_E}{\partial \tau} (\tau, {\bf u}, \vartheta) =
\sum_h E'_{h} \Phi_{E h}({\bf u}, \vartheta), \ \ 
E'_h=\left\langle \frac{\partial F_E}{\partial \tau} \Psi_{E h} \right\rangle_E ,\\\\
\ds F_L(\tau, {\bfchi}) = \sum_k L_{k} \Phi_{L k}({\bfchi}), \ \ 
L_k=\left\langle F_L \Psi_{L k} \right\rangle_L ,\\\\
\ds \frac{\partial F_L}{\partial \tau}(\tau, {\bfchi}) =
\sum_k L'_{k} \Phi_{L k}({\bfchi}), \ \ 
L'_k=\left\langle \frac{\partial F_L}{\partial \tau} \Psi_{L k} \right\rangle_L ,
\end{array}
\ \ \ \ \ \ \forall \tau \ge 0 .
\eea
Here $\Phi_{E h}({\bf u}, \vartheta)$ and $\Phi_{L k}({\bfchi})$ are eigenfunctions of
$\mathcal{L}_E$ and $\mathcal{L}_L$, respectively, while
$\Psi_{E h}({\bf u}, \vartheta)$ and $\Psi_{L k}({\bfchi})$ are eigenfunctions of the
corresponding Koopman operators $\mathcal{K}_E$ and $\mathcal{K}_L$, defined as the
adjoints of $\mathcal{L}_E$ and $\mathcal{L}_L$,
\bea
\begin{array}{l@{\hspace{-0.cm}}l}
\ds \lambda_{L k} \Phi_{L k} = \mathcal{L}_L \Phi_{L k}, \ \ \ \
\ds \lambda_{E h} \Phi_{E h} = \mathcal{L}_E \Phi_{E h}, \\\\
\ds \lambda_{L k} \Psi_{L k} = \mathcal{K}_L \Psi_{L k}, \ \ \ \
\ds \lambda_{E h} \Psi_{E h} = \mathcal{K}_E \Psi_{E h},\\\\
\ds \left\langle \Phi_{L i} \Psi_{L j} \right\rangle_L = \delta_{i j}, \ \ \ \ \
\ds \left\langle \Phi_{E i} \Psi_{E j} \right\rangle_E = \delta_{i j}, \\\\
\ds \mathcal{K}_L \equiv \mathcal{L}^\dagger_L \circ =
\nabla_{\bf x} (\circ) \cdot {\bf u}(t, {\bfchi})
= \nabla_{\bf x} (\circ) \cdot \hat{\bf u}(t, {\bfchi}), \\\\
\ds \mathcal{K}_E \equiv \mathcal{L}^\dagger_E \circ =
\nabla_{\bf u} ( \circ ) \cdot {\bf N} +
\nabla_{\bf \vartheta} ( \circ ) \cdot {\bf M} .
\end{array}
\eea
In all cases,
\bea
\begin{array}{l@{\hspace{-0.cm}}l}
\ds \Re(\lambda_{L 0})=0>
\Re(\lambda_{L 1})>
\Re(\lambda_{L 2})>\dots ,\\\\
\ds \Re(\lambda_{E 0})=0>
\Re(\lambda_{E 1})>
\Re(\lambda_{E 2})>\dots , 
\end{array}
\eea
where $\lambda_{L 0}$ and $\lambda_{E 0}$ correspond to the invariant measure.
In particular, the real parts of the eigenvalues $\lambda_{L k}$ are related to the
Lagrangian bifurcation rate $S_L\gg\Lambda_L^{(i)}>...>S_{NS}$ and to the classical
Lyapunov exponents $\Lambda_L$ ($\vert {\bfxi} \vert \rightarrow 0$) through
\bea
\ds \Re(\lambda_{L k}) \sim -\sigma^+_{L k}
- C_k S_L, \ \ \ C_k = O(1) .
\eea
where 
\bea
\ds \sigma_{L k}= \sum_{i=1}^3 \Lambda^{(i)}_{L}=0
\eea
in incompressible turbulence.

A closer inspection of Eq. (\ref{zeta_t}) reveals that the asymptotic behavior of the
correlation function $\zeta$ is entirely controlled by the spectral properties of the
Liouvillean operators $\mathcal{L}_L$ and $\mathcal{L}_E$.  
In fully developed turbulence, these operators act on distinct sets of variables and,
most importantly, possess markedly different spectral distributions.
The Lagrangian operator $\mathcal{L}_L$ is characterized by a spectrum whose real parts
are shifted far into the negative half-plane as a consequence of the high Lagrangian
bifurcation rate $S_L$, whereas the spectrum of $\mathcal{L}_E$ remains comparatively
closer to the imaginary axis.

As a result, the spectrum of the full Liouville operator
$\mathcal{L}=\mathcal{L}_L+\mathcal{L}_E$ exhibits a clear separation of time scales,
with the decay rates associated with the Lagrangian modes overwhelmingly dominating
those associated with the Eulerian dynamics.
In particular, each term in the modal expansion of $\zeta_t$ involves exponential
factors of the form
$
\exp\{(\lambda_{L k}+\lambda_{E h})t\},
$
whose real parts are bounded from above by $\Re(\lambda_{L k})$.
Since $\Re(\lambda_{L k}) \sim - C_k S_L$ with $S_L\gg \Lambda_L^{(i)}>...>S_{NS}$,
all correlation modes decay on time scales of order $S_L^{-1}$.

This establishes that the decay of $\zeta$ is governed by a genuine spectral gap
between the invariant subspace associated with the zero eigenvalue and the remainder
of the spectrum of $\mathcal{L}_L$.
Such a gap ensures an exponential and extremely rapid loss of correlation between
Eulerian and Lagrangian fields, independently of the initial condition $\zeta_0$.
Importantly, this decorrelation mechanism operates on time scales much shorter than
the Lyapunov time $1/\Lambda_L$ associated with the exponential separation of nearby
trajectories.

Hence, from a spectral-dynamical perspective, the evolution induced by $\mathcal{L}_L$
acts as a fast-mixing mechanism that continuously projects the joint PDF onto the
product measure $F_E F_L$, while suppressing all mixed Lagrangian--Eulerian modes.
The Eulerian dynamics, encoded in $\mathcal{L}_E$, evolves on significantly longer
time scales and is therefore unable to sustain persistent correlations with the
rapidly decorrelating Lagrangian field.

In the particular case of initially uncorrelated fields, $\zeta_0=0$, the above
spectral structure implies that $\zeta_t=0$ for all $t>0$, i.e., the factorization
of the joint PDF is dynamically preserved.
More generally, for arbitrary initial conditions, Lagrangian--Eulerian correlations
are destroyed exponentially fast, at a rate controlled by the Lagrangian bifurcation
 rather than by classical Lyapunov stretching.

This conclusion is consistent with the results reported in
Refs. \cite{Ottino89, Ottino90} (and references therein), which show that:
a) the equation $\dot{\bfchi}={\bf u}(t, {\bfchi})$ produces chaotic trajectories
even for relatively simple forms of ${\bf u}(t, {\bf x})$, including steady fields;
b) flows described by ${\bf u}(t, {\bfchi})$ undergo continuous and rapid stretching
and folding, resulting in an intense degree of trajectory mixing.

\bigskip
\section{Invariance of relative kinetic energy and lagrangian trajectories separation.}

Although the two representation of motion tend to rapidly decouple statistically, these are still equivalent.
In this section, a fundamental property of fully developed turbulence, which is particularly relevant for the description of the turbulent energy cascade, is highlighted within the framework of homogeneous isotropic turbulence.
This property states that the relative kinetic energy between two spatial points $\bf x$ and ${\bf x}+{\bf r}$ (Eulerian viewpoint) coincides with that between two material points $\bf X$ and ${\bf X}+{\bf r}$ (Lagrangian description), namely
\bea
\begin{array}{l@{\hspace{-0.cm}}l}
\ds \left\langle \Delta {\bf u}  \cdot  \Delta {\bf u} \right\rangle_E
=  \left\langle \Delta \dot{\bfchi} \cdot \Delta \dot{\bfchi} \right\rangle_L, \\\\
\ds \Delta {\bf u} = {\bf u}(t, {\bf x} + {\bf r})- {\bf u}(t, {\bf x}) \equiv {\bf u}'-{\bf u}, \\\\
\ds \dot{\bfxi} \equiv \Delta \dot{\bfchi} = {\bf u}(t, {\bfchi}(t, {\bf X}+{\bf r}))- {\bf u}(t, {\bfchi}(t, {\bf X}))
\equiv \dot{\bfchi}'-\dot{\bfchi}, 
\end{array}
\label{KEE}
\eea 
By invoking the assumptions of homogeneity and isotropy, Eqs.~({\ref{KEE}}) yield
\bea
\ds \left\langle \Delta u_r^2 \right\rangle_E = \left\langle \dot{\xi}_{\xi}^2 \right\rangle_L \equiv \left\langle \Lambda_L^2(r)\right\rangle_L r^2=2 u^2 \left( 1-f \right),
\label{isotropy_f2}
\eea
where
\bea
\begin{array}{l@{\hspace{-0.cm}}l}
\ds \Delta u_r = \Delta {\bf u} \cdot \frac{\bf r}{r} \equiv \left( {\bf u} (t, {\bf x}+{\bf r}) - {\bf u} (t, {\bf x})\right)  \cdot \frac{\bf r}{r}
= u_r' - u_r, \\\\
\ds \dot{\xi}_{\xi} = \dot{\bfxi}\cdot \frac{\bfxi}{\xi} \equiv \left( {\bf u} (t, {\bfchi}+{\bfxi}) - {\bf u} (t, {\bfchi})\right)  \cdot \frac{\bfxi}{\xi} = u_\xi'-u_\xi,
\end{array}
\label{Deltau xi 1}
\eea
and $f(t,r)=\langle u_r u_r' \rangle_E/u^2$ denotes the longitudinal velocity pair correlation function, while $u \equiv \sqrt{\langle {\bf u}\cdot{\bf u}\rangle_E/3}$ is defined according to Ref.~\cite{Karman38}.
It is crucial to emphasize that $\Delta u_r$ and $\dot{\xi}_{\xi}$ are two distinct quantities, which are related through Eq.~(\ref{Deltau xi 1}) but obey fundamentally different statistical laws:
$\Delta u_r$ evolves according to the Navier--Stokes equations, whereas $\dot{\xi}_{\xi}$ is governed by Lyapunov theory and by the alignment properties of $\bfxi$.
In particular, $\dot{\xi}_{\xi}$ directly reflects the exponential stretching of material line elements, which is quantified by positive Lyapunov exponents and manifests itself as an instability of the Lagrangian flow map.
From an operator-theoretic perspective, this instability, which mainly includes the contribution of lagrangian bifurcation rate ($S_L \gg \Lambda_L$), induces a separation between the zero eigenvalue associated with statistical invariants and the remainder of the spectrum of the Liouville (or equivalently Koopman) generator, giving rise to a finite spectral gap.
This spectral gap characterizes the rate at which Eulerian-Lagrangian correlations decay and sets a characteristic timescale for the irreversible transfer of kinetic energy across scales.
The stretching mechanism is therefore responsible for populating the non-zero part of the Liouville spectrum, while the folding mechanism, imposed by incompressibility and phase--space volume preservation, prevents unbounded growth and ensures the boundedness of the Liouville spectrum.
As a consequence, their statistical averages satisfy
\bea
\begin{array}{l@{\hspace{-0.cm}}l}
\ds \left\langle \dot{\xi}_{\xi} \right\rangle_L= \left\langle \Lambda_L(r)\right\rangle_L r> 0, \\\\
\ds \left\langle \Delta u_r \right\rangle_E = 0,
\end{array}
\label{Deltau xi 2}
\eea
where the first relation in (\ref{Deltau xi 2}) expresses the Lyapunov property that, in incompressible flows, neighboring trajectories continuously diverge from each other due to persistent stretching, whereas the second relation states that the mean relative velocity between two fixed spatial points vanishes in homogeneous turbulence.
Equations~(\ref{Deltau xi 2}), (\ref{isotropy_f2}) and (\ref{Deltau xi 1})  therefore establish a direct connection between the relative kinetic energy of velocity increments, the stretching--folding dynamics of material elements, and the existence of a spectral gap in the Liouville operator, identifying the latter as the operator-theoretic signature of the turbulent energy cascade.

\bigskip

\section{Range of finite--scale Lagrangian Lyapunov exponent}

Owing to the very large rate of crossings of the Jacobian--degeneracy manifold
($S_L \gg \Lambda_L$) and to fluid incompressibility, trajectories of contiguous
fluid particles diverge from each other, exhibiting a very high degree of chaos.
As a consequence, one expects a continuous distribution of both $\Lambda_L$ and
$\dot{\xi}_{\xi}$.  
The purpose of this section is to derive the limits in which $\Lambda_L$ ranges.

The evolution of the separation between trajectories is governed by
\bea
\begin{array}{l@{\hspace{-0.cm}}l}
\ds \dot{{\bfchi} } = {\bf u} (t, {\bfchi}),   \\\\
\ds \dot{{\bfxi} } = {\bf u} (t, {\bfchi}+{\bfxi}) - {\bf u} (t, {\bfchi}),
\end{array}
\label{kin finite}
\eea
where ${\bfchi}(t, {\bf X})$ and
${\bfchi}(t, {\bf X}')={\bfchi}(t, {\bf X})+{\bfxi}(t, {\bf X}', {\bf X})$
denote two trajectories associated with particles $\bf X$ and $\bf X'$, and
$\bfxi$ is their relative separation vector.
The rate of separation between trajectories is quantified by the radial component
of the relative velocity, evaluated for $\vert {\bfxi} \vert=r$, namely
\bea
\ds \dot{\xi}_{\xi} =
\frac{\dot{\bfxi} \cdot \bfxi}{\bfxi \cdot \bfxi} \ r
= \Lambda_L r .
\label{U}
\eea

Fluid incompressibility and the large Lagrangian bifurcation rate have profound
implications for both the range of variation of $\Lambda_L$ and
$\dot{\xi}_{\xi}$, as well as for their statistics.
To analyse these effects, we represent $\bfxi$ in a suitable reference frame
under the assumption of statistical isotropy:
\bea
\begin{array}{l@{\hspace{-0.cm}}l}
\ds {\bfxi} =
\sum_{k=1}^3 \xi_k {\bf e}^{(k)}
\equiv
\sum_{k=1}^3 \zeta_k \ {\mbox e}^{\varphi_k(t)} {\bf e}^{(k)}, \\\\
\ds \varphi_k(0)=0, \quad k = 1, 2, 3 .
\end{array}
\label{xi inc}
\eea
Here
$E\equiv\left( {\bf e}^{(1)}, {\bf e}^{(2)}, {\bf e}^{(3)}\right)$
is the finite--scale Lyapunov basis, which rotates with respect to the inertial
frame $\cal R$ with angular velocity $\bfomega$ determined by the local fluid
motion. The coordinates of $\bfxi$ in $E$ are
$\xi_k \equiv \zeta_k e^{\varphi_k}$.
The quantities $\zeta_k=\zeta_k(t)$ and $\varphi_k=\varphi_k(t)$,
$k=1,2,3$, are slowly varying functions of time.

Fluid incompressibility implies
\bea
\begin{array}{l@{\hspace{-0.cm}}l}
\ds \sum_{k=1}^3 \dot{\varphi}_k(t) =0,
\end{array}
\label{xi inc 1}
\eea
where $\dot{\varphi}_k(t)\equiv\Lambda_L^{(k)}$, $k=1,2,3$.
Statistical isotropy further imposes a constraint on $\varphi_k$ that is compatible
with Eq.~(\ref{xi inc 1}), which can be written as
\bea
\begin{array}{l@{\hspace{-0.cm}}l}
\ds \varphi_k(t)=\varphi(t)
\cos\!\left(\beta + \frac{2}{3}\pi (k-1)\right),
\qquad k = 1, 2, 3 .
\end{array}
\label{isotropy}
\eea

Since $\bfxi$ represents the separation between two fluid particles,
$\varphi(t)$ is a Lipschitz, monotone, differentiable function of time satisfying
\bea
\begin{array}{l@{\hspace{-0.cm}}l}
\ds \forall t \in \left(-\infty, \infty \right),
\quad 0 < \dot{\varphi}(t) \leq L < \infty, \\\\
\ds \lim_{t \rightarrow \pm \infty} \varphi(t) = \pm \infty, \\\\
\ds \lim_{t \rightarrow \pm \infty}  \inf \dot{\varphi} = 0, \\\\
\ds \lim_{t \rightarrow \pm \infty} \sup \dot{\varphi} = L < +\infty .
\end{array}
\label{lip}
\eea
Moreover, since ${\bf e}^{(1)}$ identifies the direction of maximal growth of
$\ln \vert \bfxi \vert$, $\varphi_1$ is the largest component according to
Eq.~(\ref{isotropy}), which implies $\beta=0$ for $\varphi>0$, namely
\bea
\begin{array}{l@{\hspace{-0.cm}}l}
\ds \varphi_1(t) = \varphi(t), \\\\
\ds \varphi_2(t)= \varphi_3(t)
= - \frac{\varphi(t)}{2}.
\end{array}
\label{xi inc 11}
\eea

To determine the range of variation of $\Lambda_L$, we express it in terms of
$\zeta_k$ and $\varphi_k$:
\bea
\ds \Lambda_L =
\frac{ \dot{\bfxi} \cdot \bfxi}{\bfxi \cdot \bfxi} =
\frac{\ds \sum_{k=1}^3
\left( \dot{\zeta_k} \zeta_k+ {\zeta_k}^2 \dot{\varphi}_k\right)
e^{2 \varphi_k}}
{\ds \sum_{k=1}^3 {\zeta_k}^2 e^{2 \varphi_k} } .
\label{xi lim}
\eea
The extrema of $\Lambda_L$ are obtained by taking the limits
$t \rightarrow \pm \infty$ of Eq.~(\ref{xi lim}), noting that $\varphi$ and
$\zeta_k$ are slowly varying functions and that $\varphi$ is monotone. One finds
\bea
\begin{array}{l@{\hspace{-0.cm}}l}
\ds \inf\left\lbrace \Lambda_L \right\rbrace=
\lim_{t\rightarrow - \infty} \inf \Lambda_L = -\frac{L}{2}, \\\\
\ds \sup\left\lbrace \Lambda_L \right\rbrace=
\lim_{t\rightarrow + \infty} \sup \Lambda_L = L .
\end{array}
\eea
Therefore,
\bea
\begin{array}{l@{\hspace{-0.cm}}l}
\ds \Lambda_L \in \left( -\frac{L}{2},  L \right).
\end{array}
\label{U range}
\eea
These bounds are independent of the particular realization of the Eulerian
velocity field and can equivalently be obtained by considering any specific
realization.

Furthermore, Eq.~(\ref{xi inc}) implies that $\bfxi$ tends to align with the
direction ${\bf e}^{(1)}$ of maximal growth. The relative velocity can thus be
conveniently decomposed as
\bea
\begin{array}{l@{\hspace{-0.cm}}l}
\ds \dot{\bfxi}= \left\langle \Lambda_L \right\rangle_L {\bfxi}
+ {\bfeta} + {\bfomega} \times {\bfxi},
\end{array}
\label{align}
\eea
where $\left\langle \Lambda_L \right\rangle_L {\bfxi}$ represents the contribution
of the mean Lyapunov exponent and $\bfeta$ is the residual term arising from
Eq.~(\ref{xi inc}). Accordingly,
\bea
\begin{array}{l@{\hspace{-0.cm}}l}
\ds \Lambda_L =
\left\langle \Lambda_L \right\rangle_L  +  \xi_M \cos \alpha, \\\\
\ds \mbox{where} \quad
\alpha = \widehat{{\bfxi}{\bfeta}}, \qquad
\xi_M =\frac{\vert {\bfxi}\vert \vert {\bfeta} \vert}{{\bfxi} \cdot {\bfxi}} .
\end{array}
\label{align 1}
\eea
Since $\Lambda_L \in \left( -L/2,  L \right)$ for $\alpha \in (0,\pi)$, the regime
of fully developed chaos yields
$\left\langle \Lambda_L \right\rangle_L =L/4$ and $\xi_M = 3L/4$, so that
\bea
\begin{array}{l@{\hspace{-0.cm}}l}
\ds \Lambda_L = \frac{L}{4} \left( 1 + 3 \cos \alpha \right).
\end{array}
\eea

\bigskip

\section{Statistics of finite--scale Lagrangian separation rate}

The aim of this section is to derive the distribution of $\Lambda_L$ to establish a
quantitative connection between Eqs.~(\ref{Deltau xi 2}) and (\ref{Deltau xi 1}),
and hence to relate the relative kinetic energy between two points to the energy
cascade.
The PDF of $\Lambda_L$ (or equivalently of
$\dot{\xi}_{\xi}$), denoted by $F_{\Lambda}$, can now be derived on the basis of
the preceding analysis. To this end, we start from the definition of
$\Lambda_L$,
\bea
\ds \Sigma_\Lambda: \Psi({\bfchi},\Lambda_L ) \equiv \Lambda_L -
\frac{\dot{\bfxi}\cdot {\bfxi}}{{\bfxi}\cdot {\bfxi}} =0 ,
\label{sigmax}
\eea
which defines a family of hypersurfaces
$\left\lbrace \ds \Sigma_\Lambda,\ \Lambda_L \in (-L/2, L)\right\rbrace$.
For each hypersurface $\Sigma_\Lambda$, both its measure $m(\Sigma_\Lambda)$ and
the measure of any infinitesimal element $d\Sigma_\Lambda$ are independent of
$\Lambda_L$, consistently with the hypothesis of fully developed chaos.

To compute $F_{\Lambda}$, we note that $\Lambda_L$ is not directly linked to the
Eulerian PDF $F_E(t, {\bf u}, \vartheta)$. Its PDF, which is related to the
Lagrangian distribution $F_L$, can therefore be expressed through the
Frobenius--Perron equation \cite{Nicolis95} combined with
Eq.~(\ref{sigmax}),
\bea
\begin{array}{l@{\hspace{-0.cm}}l}
\ds F_{\Lambda} ( \Lambda_L )
=
\int_{\bfchi} F_L \,
\delta\!\left( \Psi({\bfchi},\Lambda_L ) \right) \, d{\bfchi} .
\end{array}
\label{FrobeniusPerron}
\eea
Equivalently, using Eq.~(\ref{align 1}), one may write
\bea
\begin{array}{l@{\hspace{-0.cm}}l}
\ds F_\Lambda (\Lambda_L) =
\int_{\bfchi} F_L \,
\delta\!\left( \Lambda_L - \frac{L}{4}
\left(  1 +  3 \cos \alpha \right) \right) \, d{\bfchi} .
\end{array}
\label{FrobeniusPerron lyap}
\eea
Here $\delta$ denotes Dirac's delta distribution. Owing to statistical
homogeneity, $F_\Lambda$ does not depend on the spatial position $\bf x$.
Moreover, the absence of privileged directions in isotropic turbulence suggests
that $\Lambda_L$ may be uniformly distributed over its admissible interval.

This property can be demonstrated by observing that the integral in
Eq.~(\ref{FrobeniusPerron}) can be recast as a layer integral over
$\ds \Sigma_\Lambda$ \cite{Federer69},
\bea
\ds F_\Lambda( \Lambda_L)=
\int_{\Sigma_\Lambda}
\frac{F_L}{\vert {\nabla_{\bfchi} \Psi}\vert} \, d \Sigma_\Lambda .
\eea
Since $F_L$, ${\nabla_{\bfchi} \Psi}$, $d \Sigma_\Lambda$, and
$m (\Sigma_\Lambda)$ are all independent of $\Lambda_L$, it follows that
$F_\Lambda$ is constant within the interval of variation of $\Lambda_L$.
Consequently,
\bea
\ds F_\Lambda ( \Lambda_L) =
\left\lbrace
\begin{array}{l@{\hspace{-0.cm}}l}
\ds \frac{2}{3}\frac{1}{L},
\quad \mbox{if} \ \Lambda_L \in \left( -\frac{L}{2}, L\right), \\\\
\ds 0 \quad \mbox{elsewhere} .
\end{array}\right.
\label{Pl}
\eea

An alternative and equivalent derivation of Eq.~(\ref{Pl}) exploits the
statistical isotropy assumption together with
Eq.~(\ref{FrobeniusPerron lyap}).  
\begin{figure}
	\centering
	\includegraphics[width=52mm,height=70mm]{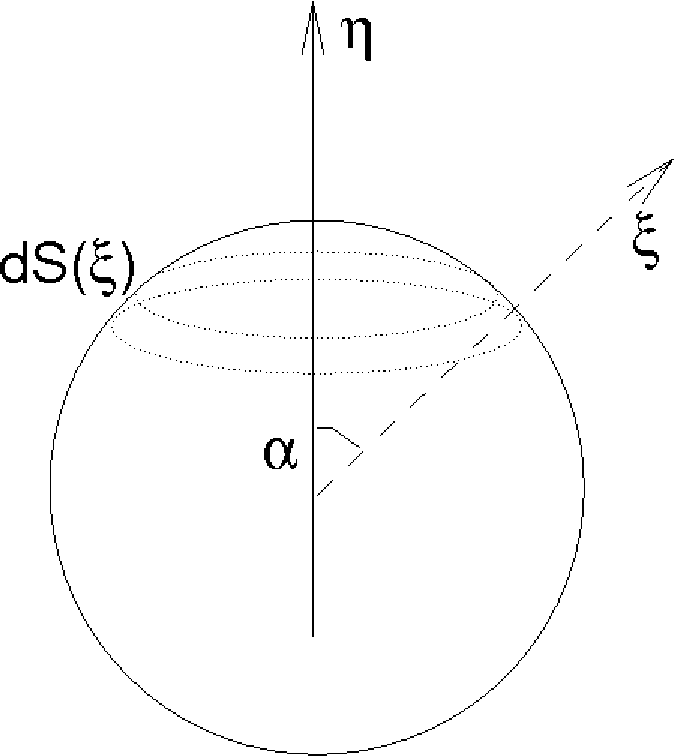}
\caption{Schematic representation of $\bfxi$ and $\bfeta$.}
\label{figura_3}
\end{figure}
Under isotropy, all orientations of $\bfxi$ are equiprobable. Therefore, the
elementary probability $F_L({\bfchi}) \, d{\bfchi}$ evaluated in the vicinity of
$\Sigma_\Lambda$ ($\Psi=0$) is restricted to those vectors $\bfxi$ such that
$\widehat{\bfxi \bfeta}\in(\alpha, \alpha+d\alpha)$. 
Hence, the elementary probability $F_L({\bfchi}) \, d{\bfchi}$ evaluated near
$\Sigma_\Lambda$ depends solely on $\alpha$ and is proportional to the elemental
surface $dS(\xi)$ shown in Fig.~\ref{figura_3},
\bea
\begin{array}{l@{\hspace{-0.cm}}l}
\ds   F_L \, d{\bfchi}
= \frac{dS(\xi)}{4 \pi \xi^2}
= \frac{2 \pi \xi^2 \sin \alpha \, d \alpha}{4 \pi \xi^2}
= -\frac{d \cos \alpha}{2},
\qquad
\alpha \in \left( 0, \pi \right).
\end{array}
\label{isotropy 1}
\eea
By combining Eqs.~(\ref{isotropy 1}), (\ref{FrobeniusPerron}) and
(\ref{align 1}), the expression for $F_\Lambda ( \Lambda_L )$ reduces to an
integral over $\cos \alpha$, with $\alpha \in (0,\pi)$, yielding
\bea
\begin{array}{l@{\hspace{-0.cm}}l}
\ds  F_\Lambda ( \Lambda_L ) =
\frac{1}{2} \int_{-1}^1
\delta \left( \Lambda_L -
\frac{L}{4} \left(  1 +  3 \cos \alpha \right) \right)
\, d \cos \alpha \\\\
\ds =
\frac{2}{3} \frac{1}{L}
\left( H \left(\Lambda_L +\frac{L}{2} \right)-
H \left( \Lambda_L  -L \right) \right),
\end{array}
\eea
where $H$ denotes the Heaviside function. It follows that $\Lambda_L$ and
$\dot{\xi}_{\xi} \equiv \Lambda_L \, r$ are uniformly distributed over their
respective intervals given by Eq.~(\ref{U range}).

This PDF implies $\langle \dot{\xi}_{\xi} \rangle_L >0$ and establishes a direct
connection between its statistical moments. In particular, the relation between
$\langle \dot{\xi}_{\xi} \rangle_L$ and
$\langle \dot{\xi}_{\xi}^2 \rangle_L$, which is instrumental for the present
analysis, allows one to express $\langle \dot{\xi}_{\xi} \rangle_L$ in terms of
the pair--velocity correlation according to Eq.~(\ref{isotropy_f2}),
\bea
\begin{array}{l@{\hspace{-0.cm}}l}
\ds \left\langle \dot{\xi}_{\xi} \right\rangle_L =
\frac{1}{2}
\sqrt{\left\langle \dot{\xi}_{\xi}^2 \right\rangle_L}
\equiv  u \sqrt{\frac{1-f}{2}} .
\end{array}
\label{mPl}
\eea
In fully developed turbulence the relative kinetic energy between
two points promotes trajectory separation. The continuous distribution of
$\Lambda_L$ arises as a consequence of the large value of
$S_L \gg \Lambda_L$ together with fluid incompressibility. The very frequent
Lagrangian bifurcations generate a continuous separation of trajectories, while
their combined effect with incompressibility determines the finite interval over
which $\Lambda_L$ varies. Trajectory divergence, associated with dynamical
instability, produces regions where $\Lambda_L>0$, whereas incompressibility
acts in the opposite sense by preserving material volume and generating regions
where $\Lambda_L<0$. Finally, isotropy eliminates privileged directions and
results in a uniform PDF over the admissible range of $\Lambda_L$.

{In conclusion, the analysis regarding the invariance of the relative kinetic energy between the Eulerian and Lagrangian frameworks leverages the fundamental property that Eulerian velocity means vanish, whereas Lagrangian means do not. This distinction arises because the latter follow a Lyapunov--based statistics, characterized by a preferential alignment along the direction of maximal growth. In a regime of fully developed chaos, a non--zero value of this relative kinetic energy effectively increases the mean separation of adjacent trajectories, resulting in a non--vanishing mean separation velocity. The analytical relations here derived strictly hold under conditions of  incompressible homogeneous and isotropic turbulence. These conditions serve as a necessary analytical framework to quantify the trajectory separation effect and, consequently, to characterize the energy cascade mechanism.
It is clear that under general non--isotropic conditions, Eq. (\ref{mPl}) cannot be expressed through a single scalar variable. However, one might empirically invoke the local isotropy of the small scales  (the Kolmogorov hypothesis), provided this remains verifiable in the specific flow configuration. We believe this idealized treatment provides a fundamental benchmark for understanding the coupling between Lagrangian dynamics and energy cascade.}

\bigskip

\section{Remark: Role of the Lagrangian bifurcation rate in the spectral gap of the Liouville generator}

The relation between finite--scale Lyapunov statistics and the spectral gap of
the Liouville generator is completed by accounting for the Lagrangian
bifurcation rate $S_L$, which represents the dominant dynamical mechanism in
fully developed turbulence.
While the finite--scale Lyapunov exponent $\Lambda_L$ characterizes local
stretching or contraction of nearby trajectories, the bifurcation rate $S_L$
measures the frequency of singular events of the Lagrangian Jacobian, producing
topological rearrangements of phase--space trajectories. In strongly chaotic
regimes one has $S_L \gg \Lambda_L$, so that bifurcations occur on time scales
much shorter than those associated with exponential separation.
Each bifurcation induces a rapid reorientation of the local stretching
directions, enhancing folding and redistributing phase--space filaments.
Accordingly, the stretching--folding cycle is primarily driven by $S_L$, while
$\Lambda_L$ governs only the stretching intensity between successive events.
As a consequence, the spectral gap of the Liouville generator,
\[
\Delta_{\mathcal{L}} \sim S_L + \langle \Lambda_L \rangle_L ,
\]
is controlled mainly by the bifurcation, with the positive mean
finite--scale Lyapunov exponent $\langle \Lambda_L \rangle_L = L/4$ providing a
secondary contribution. The bounded and uniform distribution of $\Lambda_L$
derived in Eq.~(\ref{Pl}) therefore reflects the rapid sequence of Lagrangian
bifurcations. In incompressible turbulence, the combined action of bifurcations
and incompressibility yields a finite spectral gap, ensuring fast decay of
Eulerian--Lagrangian correlations and efficient statistical relaxation.

\bigskip

\section{Closure of von K\'arm\'an--Howarth and Corrsin equations}

In this section, closures of the correlation equations are proposed on the basis of the analysis developed in the previous sections. For the reader's convenience, the von K\'arm\'an--Howarth and Corrsin equations are first recalled. These equations, obtained from the Navier--Stokes and heat equations written at two spatial points, ${\bf x}$ and ${\bf x}'={\bf x}+{\bf r}$, read
\bea
\begin{array}{l@{\hspace{-0.cm}}l}
\ds \frac{\partial f}{\partial t} = 
\ds  \frac{K}{u^2} +
\ds 2 \nu  \left(  \frac{\partial^2 f} {\partial r^2} +
\ds \frac{4}{r} \frac{\partial f}{\partial r}  \right) +\frac{10 \nu}{\lambda_T^2} f, \\\\
\ds \frac{\partial f_\theta}{\partial t} = 
\ds  \frac{G}{\theta^2} +
\ds 2 \kappa  \left(  \frac{\partial^2 f_\theta} {\partial r^2} +
\ds \frac{2}{r} \frac{\partial f_\theta}{\partial r}  \right) +\frac{12 \kappa}{\lambda_\theta^2} f_\theta,
\end{array}
\label{vk-h}
\eea
where $f=\langle u_r u_r' \rangle_E/u^2$ and $f_\theta = \langle \vartheta  \vartheta' \rangle_E/\theta^2$ are, respectively, the correlations of the radial velocity components and of the temperature; $u \equiv \sqrt{\langle u_r^2 \rangle_E}$ and $\theta \equiv \sqrt{\langle \vartheta^2 \rangle_E}$. The quantities $\lambda_T \equiv \sqrt{-1/f''(0)}$ and $\lambda_\theta \equiv \sqrt{-2/f_\theta''(0)}$ denote the Taylor and Corrsin microscales, respectively.

The functions $K$ and $G$, which account for the turbulent energy cascade, are related to $k$ and $m^*$, namely the longitudinal triple velocity correlation and the mixed triple correlation between $u_r$ and $\vartheta$, according to 
\bea
\begin{array}{l@{\hspace{+0.0cm}}l}
\ds K(r) = u^3 \left( \frac{\partial }{\partial r} + \frac{4}{r} \right) 
k(r), 
\ \ \mbox{where} \ \ 
\ds k(r) = \frac{\langle u_r^2 u_r' \rangle_{E}}{u^3}, \\\\
\ds G(r) = 2 u \theta^2 \left( \frac{\partial }{\partial r} + \frac{2}{r} 
\right) m^*(r), 
\ \ \mbox{where} \ \ 
\ds m^*(r) = \frac{\langle u_r \vartheta \vartheta' \rangle_{E}}{\theta^2 u},
\end{array}
\eea
 Equations (\ref{vk-h}) are closed once both $K$ and $G$ are expressed solely in terms of $f$ and $f_\theta$.

If these correlations were evaluated following the classical approaches \cite{Karman38, Corrsin_1, Corrsin_2}, namely as averages over the Eulerian ensemble ${\bf u} \times \vartheta$ (i.e. through $F_E(t,{\bf u},\vartheta)$), the quantities $K$ and $G$ would remain unknown unless specific assumptions on their analytical structure are introduced \cite{Karman38, Corrsin_1, Corrsin_2}. Here, instead, $K$ and $G$ are obtained by first deriving Eqs. (\ref{vk-h}) in a formal manner from the Liouville theorem, and then identifying $K$ and $G$ through the results of the previous analysis concerning the very fast exponential decay of correlations between Eulerian and Lagrangian fields.

To this end, the von K\'arm\'an--Howarth and Corrsin equations are derived by multiplying the Liouville equation (\ref{Liouville}) by $u_r u_r'$ and $\vartheta \vartheta'$, respectively, and integrating the resulting equations over $\left\lbrace {\bfchi} \right\rbrace \times\left\lbrace  \left\lbrace {\bf u}  \right\rbrace \times \left\lbrace  {\vartheta} \right\rbrace \right\rbrace $,
\bea
\begin{array}{l@{\hspace{-0.cm}}l}
\ds \int_{\bf \chi} \int_{\bf u \times \vartheta}  u_r u_r' 
\left(
 \frac{\partial F}{\partial t}-{\mathcal L}_L F - {\mathcal L}_E F
\right) 
  d {\bf u} \ d {\vartheta} d {\ds \bfchi} 
=0, \\\\
\ds \int_{\bf \chi} \int_{\bf u \times \vartheta}  \vartheta \vartheta' 
\left( 
 \frac{\partial F}{\partial t}-{\mathcal L}_L F - {\mathcal L}_E F
\right)  d {\bf u} \ d {\bf \vartheta}  d {\bfchi} 
=0,
\end{array}
\label{new}
\eea
 where the Lagrangian operator ${\mathcal L}_L$ accounts for transport terms and for the turbulent energy cascade. The contributions of ${\mathcal L}_L$ arising from Eqs. (\ref{new}) correspond to transport terms that do not modify the rates of kinetic and thermal energies \cite{Batchelor53, Corrsin_1}, and thus identify $K$ and $G$ as
 \bea
\begin{array}{l@{\hspace{+0.0cm}}l}
\ds K = \int_{\bf \chi} \int_{ \bf u \times \vartheta} \hspace{-5.mm} {\cal L}_L F \ u_r u_r' \ d {\bf u}  d {\bf \vartheta} \ d {\bfchi}=
-\int_{\bf \chi} \int_{ \bf u \times \vartheta} \hspace{-5.mm} \nabla_{\bf x} \cdot \left( F \ \dot{\bfchi}\right)  \ u_r u_r'  \ d {\bf u}  d {\bf \vartheta} \ d {\bfchi}, \\\\
\ds G = \int_{\bf \chi} \int_{\bf u \times \vartheta} \hspace{-5.mm} {\cal L}_L F \ \vartheta \vartheta'  \ d {\bf u}  d {\bf \vartheta} \ d {\bfchi}=
-\int_{\bf \chi} \int_{ \bf u \times \vartheta} \hspace{-5.mm} \nabla_{\bf x}\cdot \left( F \ \dot{\bfchi}\right) \ \vartheta \vartheta' \  d {\bf u}  d {\bf \vartheta} \ d {\bfchi}
\end{array}
\label{closures}
\eea
 The remaining terms in Eqs. (\ref{new}) reproduce the other contributions appearing in Eqs. (\ref{vk-h}). Integrating by parts Eqs. (\ref{closures}) with respect to ${\bfchi}$ and taking into account the absence of probability flux across the boundaries ($F=0$, $\forall {\bfchi} \in \partial \lbrace {\bfchi} \rbrace$), $K$ and $G$ reduce to 
\bea
\begin{array}{l@{\hspace{+0.0cm}}l}
\ds K = \int_{\bf \chi} \int_{ \bf u \times \vartheta} \hspace{-2.mm}  F 
\left( \frac{\partial u_r u_r'}{\partial {\bf x}}\cdot \dot{\bfchi}+
       \frac{\partial u_r u_r'}{\partial {\bf x}'}\cdot \dot{\bfchi}' \right)
 \  d {\bf u}  d {\bf \vartheta} \ d {\bfchi}, \\\\
\ds G = \int_{\bf \chi} \int_{ \bf u \times \vartheta} \hspace{-2.mm}  F 
\left( \frac{\partial \vartheta \vartheta'}{\partial {\bf x}}\cdot \dot{\bfchi}+
       \frac{\partial \vartheta \vartheta'}{\partial {\bf x}'}\cdot \dot{\bfchi}' \right)
 \  d {\bf u}  d {\bf \vartheta} \ d {\bfchi},
\end{array}
\label{KG (1)}
\eea
When $S_L \gg \Lambda_L^{(1)}$, the exponential decay of correlations between Lagrangian and Eulerian fields is much faster than the trajectories separation rate associated with the Lyapunov exponents; therefore $F \rightarrow F_E F_L$. Consequently, in Eq. (\ref{KG (1)}), $F_E$ acts on $u_r u_r'$ and $\vartheta \vartheta'$, whereas $F_L$ acts separately on $\dot{\bfchi}$. Moreover, in homogeneous and isotropic turbulence, the following relations hold \cite{Karman38, Corrsin_1, Corrsin_2} 
\bea
\ds \frac{\partial}{\partial {\bf x}'} \left\langle \circ \right\rangle_E= -\frac{\partial}{\partial {\bf x}} \left\langle \circ \right\rangle_E= \frac{\partial}{\partial {\bfxi}} \left\langle \circ \right\rangle_E = \frac{\partial}{\partial r} \left\langle \circ \right\rangle_E \frac{\bfxi}{\xi}
\eea
 where $\circ = u_r u_r'$, $\vartheta \vartheta'$, and $\dot{\bfxi}=\dot{\bfchi}'-\dot{\bfchi}$, with ${\bf x}'={\bfchi}(t,{\bf X}')$, ${\bf x}={\bfchi}(t,{\bf X})$, $\dot{\bfchi}'\equiv \dot{\bfchi}(t,{\bf X}')$, and $\dot{\bfchi}\equiv \dot{\bfchi}(t,{\bf X})$. Accordingly, $K$ and $G$ become
\bea
\begin{array}{l@{\hspace{+0.0cm}}l}
\ds K(r) =
u^2 \frac{\partial f }{\partial r}  \left\langle \dot{\xi}_{\xi} \right\rangle_L =
 u^3 \sqrt{\frac{1-f}{2}} \ \frac{\partial f}{\partial r}, \\\\
\ds G(r) = \theta^2
\frac{\partial f_\theta}{\partial r} \left\langle \dot{\xi}_{\xi} \right\rangle_L=
 u \theta^2 \sqrt{\frac{1-f}{2}} \ \frac{\partial f_\theta}{\partial r},
\end{array}
\label{K}
\label{K closure}
\eea
These expressions do not involve second--order derivatives of the autocorrelations; hence, Eqs. (\ref{K}) are not closures of diffusive type. Rather, they arise from the divergence of contiguous trajectories in fully developed turbulence, which is in turn the consequence of a bifurcation rate much larger than the Lyapunov exponents. According to Eqs. (\ref{mPl}), the closures (\ref{K closure}) show how the mean relative kinetic energy between two points produces the separation trajectory which drives the energy cascade phenomenon. This latter can be interpreted as a propagation mechanism across the scales $r$, occurring with a propagation speed $\ds \left\langle \dot{\xi}_\xi\right\rangle_L$ that depends on $r$ and $u$.

The main advantage of Eqs. (\ref{K}) with respect to other closures is that they do not rely on phenomenological assumptions. They are derived from the statistical independence of $\bfxi$ and ${\bf u}$, which follows from the rapid exponential decorrelation between Eulerian and Lagrangian fields, much faster than that associated with Lyapunov exponents. This makes it possible to express $K$ and $G$ analytically by separating the effects of trajectories divergence (Lagrangian contribution) from those of velocity field fluctuations (Eulerian contribution). Owing to their theoretical foundation, Eqs. (\ref{K closure}) do not involve free model parameters.

These closures coincide with those previously obtained by the author in Refs. \cite{deDivitiis_1, deDivitiis_4, deDivitiis_5, deDivitiis_8}, where the formulas were derived under statistical ergodicity assumptions concerning trajectories separation and bifurcation rates. In the present framework, however, this result acquires a deeper dynamical meaning through the spectral properties of the Liouville operator. Here, in contrast, Eqs. (\ref{K closure}) are obtained by means of the Liouville--Koopman spectral analysis within the framework of fully developed chaos. In particular, the existence of a finite spectral gap of the Liouville operator associated with the Navier--Stokes dynamics plays a central role. This gap, generated by the extremely rapid sequence of Lagrangian bifurcations, induces an exponential decay of correlations between Eulerian observables and Lagrangian trajectories, thereby justifying the factorization $F \rightarrow F_E F_L$ and the ensuing statistical independence.

The novelty of the present work with respect to Refs. \cite{deDivitiis_1, deDivitiis_4, deDivitiis_5, deDivitiis_8, deDivitiis_9, deDivitiis_10} lies in the explicit interpretation of the energy cascade in terms of the spectral gap of the Liouville operator associated with the Navier--Stokes equations. This gap arises directly from Lagrangian bifurcation rates that are much larger than the Lyapunov exponents, and leads to statistical independence between Eulerian and Lagrangian fields. A further novel aspect is the invariance of the relative kinetic energy between two points, shown to be the fundamental driver of the separation rate of contiguous trajectories. This separation rate is controlled by the Lagrangian bifurcation rather than by the Lyapunov exponents.

For the results obtained using Eqs. (\ref{K}), the reader is referred to the data reported in Refs. \cite{deDivitiis_1, deDivitiis_2, deDivitiis_4, deDivitiis_5, deDivitiis_8}. In particular, Refs. \cite{deDivitiis_1, deDivitiis_4, deDivitiis_5, deDivitiis_2} demonstrate that Eqs. (\ref{K}) provide an accurate description of the energy cascade, reproducing negative values of the skewness of velocity and temperature increments 
\bea
\begin{array}{l@{\hspace{+0.0cm}}l}
\ds H_{u 3}(r) \equiv 
\frac{\langle (\Delta u_r)^3 \rangle }{\langle (\Delta u_r)^2 \rangle^{3/2}} 
=
\frac{6 k(r)}{(2(1-f(r)))^{3/2}} \\\\
\ds H_{\theta 3}(r) \equiv 
\frac{\langle (\Delta \vartheta)^2 \Delta u_r \rangle }
{\langle (\Delta \vartheta)^2 \rangle {\langle (\Delta u_r)^2 \rangle}^{1/2}}=
\frac{4 m^*}{ 2(1-f_\theta(r)) (2(1-f(r)))^{1/2}} 
\end{array}
\label{H3}
\eea
 and, in particular,
\bea
\begin{array}{l@{\hspace{+0.0cm}}l}
\ds H_{u 3}(0) = \lim_{r \rightarrow 0} H_{u 3}(r) = - \frac{3}{7}, \\\\
\ds H_{\theta 3}(0) = \lim_{r \rightarrow 0} H_{\theta 3}(r) = - \frac{1}{5}, 
\end{array}
\eea
 in agreement with the literature \cite{Chen92, Orszag72, Panda89, Anderson99, Carati95, Kang2003}, with the Kolmogorov law, and with temperature spectra consistent with the theoretical arguments of Kolmogorov, Obukhov--Corrsin, and Batchelor \cite{Batchelor_2, Batchelor_3, Obukhov}, as well as with experimental \cite{Gibson, Mydlarski} and numerical \cite{Rogallo, Donzis} results.

On the basis of these closures, we now provide approximate estimates of $C_2$, $C_\theta$, and $C_B$, namely the Kolmogorov, Corrsin--Obukhov, and Batchelor constants, respectively, implicitly defined by \bea
\begin{array}{l@{\hspace{+0.0cm}}l}
\ds f \approx 1-\frac{C_2}{2 \ u^2} \left( \varepsilon r \right)^{2/3}, \ \mbox{inertial range} \\\\
\ds f_\theta \approx 1-\frac{C_\theta}{2 \ \theta^2} \ \varepsilon^{-1/3} \varepsilon_\theta \ r^{2/3}, \ \mbox{inertial--convective range} \\\\
\ds f \approx 1- \frac{1}{2} \left( \frac{r}{\lambda_T}\right)^2 , \ \mbox{viscous range} \\\\
\ds f_\theta \approx 1-\frac{C_B}{2 \ \theta^2} \ \frac{\varepsilon_\theta}{q} \ \ln\left( \frac{r}{l_B}\right), \ \mbox {viscous--convective range} 
\end{array}
\label{constants}
\eea
 where $\varepsilon_\theta=2 \kappa \left\langle \nabla_{\bf x} \vartheta \cdot \nabla_{\bf x} \vartheta \right\rangle_E$, $q =\sqrt{\varepsilon/\nu}$, and $\ell_B=\ell_K/Pr$, with $\ell_B$, $\ell_K$, and $Pr$ denoting the Batchelor scale, Kolmogorov scale, and Prandtl number, respectively. Approximate estimates of these constants are obtained by substituting Eqs. (\ref{constants}) and the proposed closure (\ref{K}) into Eqs. (\ref{vk-h}), and by equating the terms of same order of Eqs. (\ref{vk-h}), yielding 
\bea
 \begin{array}{l@{\hspace{+0.0cm}}l}
 \ds C_2 = 4^{2/3} \simeq 2.519, \\\\
 \ds C_\theta = \frac{6}{\sqrt{C_2}} \simeq 3.779, \\\\
 \ds C_B = 4\sqrt{15} \simeq 15.49 
\end{array} 
\eea
accordingly, using Eq.(\ref{constants}) and the proposed closures, 4/5 Kolmogorov law and Yaglom relation are obtained in the inertial and inertial--convective subranges, namely
\bea
 \begin{array}{l@{\hspace{+0.0cm}}l}
\ds \left\langle \left( \Delta u_r \right)^3 \right\rangle_E = -\frac{4}{5} \ {\varepsilon \ r}, \\\\
\ds \left\langle \Delta u_r (\Delta \vartheta)^2 \right\rangle_E  = -\frac{2}{3} \ {\varepsilon_\theta r}, 
 \end{array}
\eea
More detailed results are reported in Refs.~\cite{deDivitiis_1, deDivitiis_4, deDivitiis_8, deDivitiis_2}, where it is shown that the proposed closure~(\ref{K}) reproduces the Kolmogorov law in the inertial range with a Kolmogorov constant of order 2, and yields distinct scaling laws for the temperature correlation depending on $Pr$ and $R_T$, in agreement with established results in the literature. 
These closures generate self--similar correlation functions over an appropriate range of separations $r$, consistently with the statistical decoupling of Eulerian and Lagrangian descriptions established in this work. 
Such self--similarity should therefore be regarded as a robust signature of bifurcation--driven dynamics: the persistent divergence of fluid--particle trajectories induced by Lagrangian bifurcations, whose characteristic rate substantially exceeds that associated with Lyapunov exponents, sets the dominant timescale of decorrelation and ultimately controls the observed scaling behavior.

We conclude by discussing the limitations of the proposed closures. These limitations stem directly from the hypotheses underlying Eqs. (\ref{K}): they are valid only in regimes of fully developed chaos, where turbulence exhibits homogeneity and isotropy. During transitional stages of turbulence, or in more complex configurations involving specific boundary conditions, such as the presence of walls, Eqs. (\ref{K}) are no longer applicable. However, even in the absence of statistical homogeneity and isotropy, the energy cascade can still be identified through ${\cal L}_L$ and the Liouville theorem, whereas the determination of separation-rate statistics, the closure of correlation equations, and other correlations such as pressure--velocity correlations remain highly challenging problems, strongly dependent on the specific physical configuration.

\bigskip

\section{Conclusion \label{Conclusion}}

In this article, we have established a rigorous theoretical framework explaining why, despite their formal equivalence, Eulerian and Lagrangian descriptions of motion become statistically independent in fully developed homogeneous and isotropic turbulence. By invoking Liouville theorem and analyzing the structure of the associated generator, we have shown that the decay of Eulerian-–Lagrangian correlations is controlled by a spectral gap whose magnitude is primarily determined by the bifurcation rate of the velocity gradient. This rate, which reflects the continuous topological rearrangements of the phase-–space flow induced by the Navier–-Stokes dynamics, dominates over the contribution arising from Lyapunov exponents, thereby clarifying the distinct physical roles played by bifurcation mechanisms and dynamical instability.

The exponential relaxation of the joint PDF toward a factorized product of Eulerian and Lagrangian marginals emerges as a robust and universal property in developed turbulence, valid for arbitrary initial conditions. Once factorization is achieved --or if it is imposed initially-- it is preserved by the dynamics, with each marginal evolving independently according to its own representation. This result provides a precise probabilistic meaning to the statistical decoupling of Eulerian and Lagrangian turbulence and identifies the bifurcation rate as the key timescale governing this mechanism.

We have further shown that the formal equivalence between the two descriptions entails the invariance of the relative kinetic energy between arbitrary points. In conjunction with the intrinsic asymmetry of finite–-scale Lyapunov exponent distributions in incompressible flows, this invariance yields a coherent and quantitative interpretation of particle-–pair dispersion and of the turbulent energy cascade as a propagation across scales rather than as a purely diffusive phenomenon.

Finally, the present analysis naturally leads to nondiffusive closure relations for the von K\'arm\'an-–Howarth and Corrsin equations. The fact that these closures coincide with those previously derived by the author provides strong theoretical support for their validity and underscores the central role of bifurcation-–driven dynamics and Liouville spectral properties in the statistical theory of turbulence.

\bigskip 

\section{Acknowledgments}

This work was partially supported by the Italian Ministry for the Universities 
and Scientific and Technological Research (MIUR). 

\bigskip

\end{document}